\documentclass[10pt,journal,compsoc]{IEEEtran}
\ifCLASSOPTIONcompsoc
  \usepackage[nocompress]{cite}
\else
  \usepackage{cite}
\fi
\ifCLASSINFOpdf
\else
\fi
\DeclareMathAlphabet{\mathsfbr}{OT1}{cmss}{m}{n}
\SetMathAlphabet{\mathsfbr}{bold}{OT1}{cmss}{bx}{n}
\DeclareRobustCommand{\msf}[1]{%
  \ifcat\noexpand#1\relax\msfgreek{#1}\else\mathsfbr{#1}\fi
}

\makeatletter
\newcommand{\msfgreek}[1]{\csname s\expandafter\@gobble\string#1\endcsname}
\makeatother

\DeclareFontEncoding{LGR}{}{} 
\DeclareSymbolFont{sfgreek}{LGR}{cmss}{m}{n}
\SetSymbolFont{sfgreek}{bold}{LGR}{cmss}{bx}{n}
\DeclareMathSymbol{\salpha}{\mathord}{sfgreek}{`a}
\DeclareMathSymbol{\sbeta}{\mathord}{sfgreek}{`b}
\DeclareMathSymbol{\sgamma}{\mathord}{sfgreek}{`g}
\DeclareMathSymbol{\sdelta}{\mathord}{sfgreek}{`d}
\DeclareMathSymbol{\sepsilon}{\mathord}{sfgreek}{`e}
\DeclareMathSymbol{\szeta}{\mathord}{sfgreek}{`z}
\DeclareMathSymbol{\seta}{\mathord}{sfgreek}{`h}
\DeclareMathSymbol{\stheta}{\mathord}{sfgreek}{`j}
\DeclareMathSymbol{\siota}{\mathord}{sfgreek}{`i}
\DeclareMathSymbol{\skappa}{\mathord}{sfgreek}{`k}
\DeclareMathSymbol{\slambda}{\mathord}{sfgreek}{`l}
\DeclareMathSymbol{\smu}{\mathord}{sfgreek}{`m}
\DeclareMathSymbol{\snu}{\mathord}{sfgreek}{`n}
\DeclareMathSymbol{\sxi}{\mathord}{sfgreek}{`x}
\DeclareMathSymbol{\somicron}{\mathord}{sfgreek}{`o}
\DeclareMathSymbol{\spi}{\mathord}{sfgreek}{`p}
\DeclareMathSymbol{\srho}{\mathord}{sfgreek}{`r}
\DeclareMathSymbol{\ssigma}{\mathord}{sfgreek}{`s}
\DeclareMathSymbol{\stau}{\mathord}{sfgreek}{`t}
\DeclareMathSymbol{\supsilon}{\mathord}{sfgreek}{`u}
\DeclareMathSymbol{\sphi}{\mathord}{sfgreek}{`f}
\DeclareMathSymbol{\schi}{\mathord}{sfgreek}{`q}
\DeclareMathSymbol{\spsi}{\mathord}{sfgreek}{`y}
\DeclareMathSymbol{\somega}{\mathord}{sfgreek}{`w}

\DeclareMathSymbol{\svarsigma}{\mathord}{sfgreek}{`c}

\DeclareMathSymbol{\sGamma}{\mathalpha}{sfgreek}{`G}
\DeclareMathSymbol{\sDelta}{\mathalpha}{sfgreek}{`D}
\DeclareMathSymbol{\sTheta}{\mathalpha}{sfgreek}{`J}
\DeclareMathSymbol{\sLambda}{\mathalpha}{sfgreek}{`L}
\DeclareMathSymbol{\sXi}{\mathalpha}{sfgreek}{`X}
\DeclareMathSymbol{\sPi}{\mathalpha}{sfgreek}{`P}
\DeclareMathSymbol{\sSigma}{\mathalpha}{sfgreek}{`S}
\DeclareMathSymbol{\sUpsilon}{\mathalpha}{sfgreek}{`U}
\DeclareMathSymbol{\sPhi}{\mathalpha}{sfgreek}{`F}
\DeclareMathSymbol{\sPsi}{\mathalpha}{sfgreek}{`Y}
\DeclareMathSymbol{\sOmega}{\mathalpha}{sfgreek}{`W}

\DeclareRobustCommand{\mcal}[1]{%
  \ifcat\noexpand#1\relax\mathnormal{#1}\else\cal{#1}\fi
}
\DeclareRobustCommand{\BM}[1]{%
  \ifcat\noexpand#1\relax\bm{\boldUppercaseItalicGreek{#1}}\else\bm{#1}\fi
}
\makeatletter
\newcommand{\boldUppercaseItalicGreek}[1]{\csname var\expandafter\@gobble\string#1\endcsname}
\makeatother

\newcommand{\V}[1]{\bm{#1}} 
\newcommand{\Set}[1]{{\mcal{#1}}} 

\newcommand{\E}[1]{\mathbb{E}\left\{#1\right\}}

\usepackage{bm}
\usepackage{cite}
\usepackage{color}
\usepackage{psfrag}
\usepackage{acronym}
\usepackage{amsmath}
\usepackage{amssymb}
\usepackage{amsfonts}
\usepackage[colorlinks = true, allcolors = black]{hyperref}
\usepackage{latexsym}
\usepackage{mathrsfs}
\usepackage{epstopdf}
\usepackage{algorithm}
\usepackage{mathtools}
\usepackage[caption=false, font=scriptsize]{subfig}
\usepackage{algorithmic}
\usepackage{relsize}
\usepackage{comment}

\graphicspath{{figures/}}

\DeclareMathOperator*{\argmax}{arg\,max}
\DeclareMathOperator*{\argmin}{arg\,min}

\newcommand{\st}{\operatorname{s.t.}}

\newtheorem{definition}{Definition}
\newtheorem{lemma}{Lemma}
\newtheorem{proposition}{Proposition}
\newtheorem{theorem}{Theorem}

\newtheorem{remark}{Remark}

\definecolor{green}{rgb}{0, 0.5, 0}
\definecolor{pink}{rgb}{1, 0, 1}

\acrodef{agi}[AgI]{information}
\acrodef{alg}[ALG]{augmented layered graph}
\acrodef{sfc}[SFC]{service function chain}
\acrodef{mec}[MEC]{multi-access edge computing}
\acrodef{ER}[ER]{efficient route}

\begin{document}
%
\title{
    Joint Compute-Caching-Communication Control for Online Data-Intensive Service Delivery 
}
%
%
%
%

\author{Yang~Cai,~\IEEEmembership{Student~Member,~IEEE},
        Jaime~Llorca,~\IEEEmembership{Member,~IEEE},
        Antonia~M.~Tulino,~\IEEEmembership{Fellow,~IEEE}, and
        Andreas~F.~Molisch,~\IEEEmembership{Fellow,~IEEE}
\IEEEcompsocitemizethanks{\IEEEcompsocthanksitem Part of this work will be submitted to IEEE GlobeCom 2022 \cite{cai2022CCC_arXiv2}.
\IEEEcompsocthanksitem Y. Cai and A. F. Molisch are with the Department of Electrical Engineering, University of Southern California, Los Angeles, CA 90089, USA.\protect\\
E-mail: yangcai@usc.edu; molisch@usc.edu
\IEEEcompsocthanksitem J. Llorca is with New York University, NY 10012, USA.\protect\\
E-mail: jllorca@nyu.edu
\IEEEcompsocthanksitem A. M. Tulino is with New York University, NY 10012, USA, and also with the University\`{a} degli Studi di Napoli Federico II, Naples 80138, Italy.\protect\\
E-mail: atulino@nyu.edu; antoniamaria.tulino@unina.it
\IEEEcompsocthanksitem This work was supported by the National Science Foundation (NSF) under CNS-1816699.}
}

\IEEEtitleabstractindextext{
\begin{abstract}
Emerging Metaverse applications, designed to deliver highly interactive and immersive experiences that seamlessly blend physical reality and digital virtuality, are accelerating the need for distributed compute platforms with unprecedented storage, computation, and communication requirements. To this end, the integrated evolution of next-generation networks (e.g., 5G and beyond) and distributed cloud technologies (e.g., fog and mobile edge computing), have emerged as a promising paradigm to address the interaction- and resource-intensive nature of Metaverse applications. In this paper, we focus on the design of control policies for the joint orchestration of compute, caching, and communication (3C) resources in next-generation distributed cloud networks for the efficient delivery of Metaverse applications that require the real-time aggregation, processing, and distribution of multiple live media streams and pre-stored digital assets. We describe Metaverse applications via directed acyclic graphs able to model the combination of real-time stream-processing and content distribution pipelines. We design the first throughput-optimal control policy that coordinates joint decisions around (i) routing paths and processing locations for live data streams, together with (ii) cache selection and distribution paths for associated data objects. We then extend the proposed solution to include a max-throughput database placement policy and two efficient replacement policies. In addition, we characterize the network stability regions for all studied scenarios. Numerical results demonstrate the superior performance obtained via the novel multi-pipeline flow control and 3C resource orchestration mechanisms of the proposed  policy, compared with state-of-the-art algorithms that lack full 3C integrated control.
\end{abstract}

\begin{IEEEkeywords}
Metaverse, virtual/augmented reality, distributed cloud, mobile edge computing, caching, network control, real-time stream processing.
\end{IEEEkeywords}}

\maketitle

\IEEEpubid{
\begin{minipage}{3 \columnwidth}
    \centering
    {\scriptsize
    This work has been submitted to the {IEEE} for possible publication. Copyright may be transferred without notice, after which this version may no longer be accessible.
    }
\end{minipage}
}

\IEEEdisplaynontitleabstractindextext

%
\IEEEpeerreviewmaketitle

\IEEEraisesectionheading{\section{Introduction}\label{sec:introduction}}

%
%
%
%

\IEEEPARstart{A}{ugmented} \ac{agi} services, referring to a wide range of applications that deliver information of real-time relevance resulting from online aggregation, processing and distribution of source media streams, such as system automation (e.g., smart homes/factories/cities, self-driving cars) and metaverse experiences (e.g., multiplayer gaming, immersive video, virtual/augmented reality), are driving unprecedented requirements for communication, computation, and storage resources \cite{cai2022metaverse}.
To address this need, \ac{mec} has emerged as a promising solution, which facilitates users' access to nearby computation resources \cite{mach17mecsurvey}.
With continued advances in network virtualization and programmability \cite{SDN_NFV13}, the resulting {\em distributed cloud/computing network} architecture allows flexible and elastic deployment of next-generation disaggregated services composed of multiple software functions, which can be dynamically instantiated and executed at different network locations.

In addition to the {\em interaction-} and {\em compute-intensive} nature, an increasingly relevant feature exhibited by \ac{agi} services, especially the Metaverse applications, is {\em data-intensive}.
In the Metaverse, user experiences result from real-time aggregation, processing, and distribution of {\em multiple} live media streams (collected by sensors) and digital assets (pre-stored in network). Augmented reality (AR), as illustrated in Fig. \ref{fig:service}, is a clear example, which enriches initial videos with scene objects to generate enhanced experiences that can be consumed by users \cite{sun2019mobileVR}. Face/object recognition is another example, where the access to training samples is required to classify images uploaded by users \cite{xu2018face}. 

Indeed, the efficient delivery of data-intensive services requires joint orchestration of computation, caching and communication (3C) resources and end-to-end optimization of the associated decisions, including the following aspects:
(i) {\bf\em packet processing:} where to execute the service functions in order to process the multiple source data streams, (ii) {\bf\em packet routing:} how to route each data stream to the processing location, and (iii) {\bf\em packet caching:} where to create the digital objects, and how to place and replace them in the network.
In addition, the processing, routing and caching problems must be addressed in an online manner, in response to stochastic network conditions and service demands.

\subsection{Related Works}

\subsubsection{Computation and Communication}

The distributed cloud network control problem has been intensely studied in the recent literature, mainly focusing on virtual network function (VNF) placement and service embedding (i.e., flow routing), 
especially for applications that can be modeled by \acp{sfc}, with the single task offloading problem \cite{CheHao:J18} as a special case \cite{cai2020mec}.

One line of work studies this problem in a {\em static} setting, with the objective of either maximizing accepted service requests \cite{huang2021throughput,yue2021throughput}, or minimizing overall operational cost \cite{barcelo2016IoT,yue2021resource}. While useful for long timescale end-to-end service optimization, these solutions exhibit two main limitations: first, the problem is formulated as a {\em static} optimization problem without considering the dynamic nature of service requests, a critical aspect in next-generation \ac{agi} services; second, due to the combinatorial nature of the problem, the corresponding formulations typically take the form of (NP-hard) mixed integer programs (MIP), and either heuristic solutions or loose approximation algorithms are developed, compromising the quality of the resulting solution.

To address the \ac{sfc} optimization problem in dynamic scenarios, one needs to make joint packet processing and routing decisions in an online manner, as well as traffic scheduling and resource allocations, in response to stochastic system states (e.g., service demands, link capacities). Among existing techniques, Lyapunov drift control, firstly applied to communication networks \cite{TasEph:J92,Nee:B10,SinMod:J18}, proves to be a powerful tool to guide the design of throughput-optimal cloud network control policies, such as DCNC \cite{FenLloTulMol:J18a} and UCNC \cite{zhang2021multicast}, by dynamically exploring processing/routing diversity. In general, centralized routing approaches, e.g., UCNC, can attain better delay performance than distributed routing, e.g., DCNC, by delivering data packets along acyclic paths \cite{cai2022multicast_arxiv,cai2021multicast}.

\subsubsection{Caching and Communication}

Over the past decade, the dramatic growth of user demands for multimedia contents (e.g., videos) has fueled rapid advances in caching techniques, especially at the wireless edge. By storing copies of popular contents close to users, the network traffic and latency for content retrieval and distribution are significantly reduced \cite{ji16basic,shanmugam2013femtocaching,golrezaei2013d2d}.

Caching policy and transmission policy are two key elements in content distribution network design, dealing with (i) content placement in the network, and (ii) content delivery to users, respectively. Driven by different performance metrics, various caching policies have been designed, e.g., throughput \cite{cai2019multilink}, delay \cite{shanmugam2013femtocaching}, energy efficiency \cite{gregori2016caching}, etc., and some of them are jointly optimized with the employed communication techniques, e.g., non-orthogonal multiple access (NOMA) \cite{ding2018noma}, multiple-input and multiple-output (MIMO) \cite{liu2015mimo}, coded-multicast \cite{Maddah-Ali2016coded}, etc..

In multi-hop networks, flow routing, including the selections of the caching node to create and the path to deliver the content, plays an important role in the content distribution. In addition, the joint optimization of caching and routing can benefit the overall network performance \cite{paschos2018caching}. Some existing studies propose formulations targeting either throughput maximization \cite{liu2019JCR}, or service cost minimization \cite{ioannidis2018JCR}, and approximation algorithms are developed to address the resulting MIP problems.

\subsubsection{Joint 3C Optimization}

While there is a large body of works on the integration of computation-communication and caching-communication technologies into network design, the integration of 3C is a less explored topic.

Two combinations, computing-assisted information centric networking (ICN) and cache-enabled \ac{mec}, are studied in \cite{zhou2018CCC}, and both of them are promising directions for 3C integration. In cache-enabled \ac{mec}, one topic is service caching/provisioning, dealing with service functions/programs with non-trivial storage requirements \cite{xu2018service}; another topic is data caching, i.e., caching frequently used databases \cite{mao17mecsuvery}, such as processed results (of previous tasks) that might be repeatedly requested \cite{ndikumana2020CCC}, which can save the latency and computation resources to generate the content.

In this paper, we focus on the data caching aspect in the delivery of {\em data-intensive services} (in particular, Metaverse applications), assuming that the cached digital objects (or static objects) will be aggregated with user-specific data (or live data) and processed to generate highly personalized experiences for different users. Under such assumption, \cite{poularakis2020approximation} develops approximation algorithms for the data-intensive service embedding problem in a static setting; the dynamic (but simplified) setting is investigated in \cite{yeh2021deco}, focusing on the digital object retrieval and processing while ignoring the live service chain routing pipeline.

\begin{figure}
    \centering
    \includegraphics[width = .99 \columnwidth]{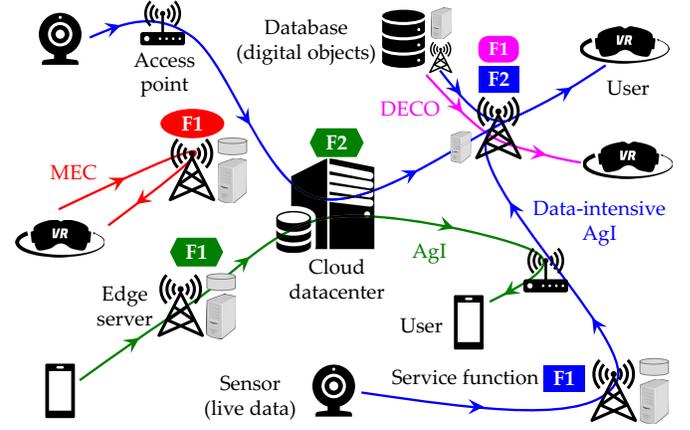}
    \caption{Network and service models studied in this paper and related works.
    Data-intensive \ac{agi} \cite{poularakis2020approximation} (this paper): distributed cloud network, service DAG (see Section \ref{sec:service_dag}) with both live data and digital objects.
    DECO \cite{yeh2021deco}: distributed cloud network, one-step processing with digital objects.
    \ac{mec} \cite{CheHao:J18}: single computing server, one-step processing with live data.
    \ac{agi} \cite{FenLloTulMol:J18a,zhang2021multicast}: distributed cloud network, \ac{sfc} with live data.
    }
    \label{fig:scenario}
\end{figure}

\subsection{Our Contributions}
\label{sec:contributions}

In this paper, we investigate the problem of {\em joint 3C control for online data-intensive service delivery}. As illustrated in Fig. \ref{fig:scenario}, the data-intensive service can include multiple functions, each requiring multiple source data streams as inputs that can be live data (collected by sensors), static objects (pre-stored in network), or intermediate results (generated by previous processing), e.g., F2 in blue.

Compared to existing service models, e.g., MEC, DECO, AgI as illustrated in Fig. \ref{fig:scenario}, two challenges arise in the delivery of data-intensive AgI services:
(i) processing location selection affects not only the processing procedure, but also the transmission loads of {\em all} input streams,
(ii) the digital object input can be created (by replication) at {\em any} caching location in an {\em on-demand} manner (i.e., per service function's request), which is fundamentally different from a live data input generated by the end user, with a {\em fixed} source and a {\em spontaneous} rate. While existing cloud network control policies \cite{CheHao:J18,FenLloTulMol:J18a,zhang2021multicast,yeh2021deco}, designed for simplified service models, are not able to handle these challenges, not to mention the coupling of them, i.e., jointly selecting the processing and caching locations. We term this problem {\em multi-pipeline flow control}.

Another key element that impacts the service delivery performance is the caching policy design, including ``which databases to cache'' and ``at which node''. Similar to existing works in caching-communication integration, database placement should be jointly optimized with the flow control decisions, but going beyond flow routing to also include flow processing, especially for heterogeneous networks with highly-distributed 3C resources. Furthermore, when considering time-varying popularity distribution (or service demand statistics), the service delivery performance shall benefit from the dynamic adjustment of the caching policy. These problems are collectively referred to as {\em joint 3C resource orchestration}.

In this paper, we address the above two problems, and our contributions can be summarized as follows:
\begin{enumerate}
    
    \item We characterize the network stability regions for data-intensive \ac{agi} service delivery, in the settings of (i) fixed and (ii) dynamic database placements.
    
    \item We design the first throughput-optimal control policy, DI-DCNC,  that coordinates joint decisions around (i) routing paths and processing locations for live data streams, together with (ii) cache selection and distribution paths for associated static data objects, under any fixed database placement.
    
    \item We propose a database placement policy targeting throughput maximization, and derive an equivalent mixed integer linear programming (MILP) problem that can be used for implementation.
    
    \item We develop two database replacement policies able to adapt to time-varying demand statistics, based on online estimations of (i) popularity distribution, and (ii) database score, respectively.
    
\end{enumerate}

The rest of this paper is organized as follows. In Section \ref{sec:system_model}, we introduce the models for cache-enabled edge cloud and data-intensive service. In Section \ref{sec:policy_space}, we define the policy space and characterize the network stability regions. We devise the DI-DCNC control policy (for fixed database placement) in Section \ref{sec:DI_DCNC}, followed by a max-throughput database placement policy in Section \ref{sec:placement}, as well as two replacement policies in Section \ref{sec:replacement}. Section \ref{sec:experiments} present the numerical results, and conclusions are drawn in Section \ref{sec:conclusions}.

Frequently used notations are summarized in Table \ref{tbl:usedsymbols}.

\begin{table}[t]\label{table1}
\footnotesize  
\newcommand{\tabincell}[2]{\begin{tabular}{@{}#1@{}}#2\end{tabular}}
\caption{Table of Notations} \vspace{-10pt}
\centering
\renewcommand{\arraystretch}{1.2}%
\begin{tabular}{p{1.5cm} p{6.5cm}}  
\hline
Symbol & Description \\
\hline
$\Set{V}, \Set{E}$ & Node and link sets of the actual network. \\
$C_i, C_{ij}, S_i$ & Processing/transmission/storage resources. \\
$\phi$ & (data-intensive) AgI service. \\ 
$\Set{K}, F_k$ & Set of databases, the size of database $k$. \\
$(\xi, r, k, \zeta)$ & Scaling factor, workload, static object, merging ratio. \\
$m$ & Processing stage (and function index). \\
$s, \Set{V}(k), d$ & Live source, set of static sources, destination. \\
$a^{(c)}(t), \lambda^{(c)}$ & Arrivals of client $c$, arrival rate. \\
$\Set{V}^{(\phi)}, \Set{E}^{(\phi)}$ & Node and edge sets of ALG for service $\phi$. \\
$o_m'$ & Super static source (of stage $m$ static objects). \\
$\sigma, \Set{F}_c(x)$ & Efficient route and the set of them. \\
$a^{(c, \sigma)}(t)$ & Number of service requests selecting ER $\sigma$. \\
$\theta^{(m)}$ & Processing location for function $m$. \\
$\Lambda(x), \Lambda$ & Stability region (under placement $x$/replacement). \\
$\tilde{Q}(t), Q(t)$  & Virtual queue and normalized virtual queue. \\
$p^{(c)}$  & Popularity distribution of service requests. \\
$x_{i, k}$ & Caching variable (if database $k$ is cached at node $i$). \\
$f_{\imath \jmath}^{(c)}, f'^{(k)}_{ij}$ & Live/static flows in ALG/actual network. \\
$w_{\imath \jmath}^{(c)}, \rho_{ij}^{(c, \sigma)}$ & Resource load on an ALG edge/actual link. \\
$V_{i, k}$ & Score (of database $k$ at node $i$). \\
\hline
\end{tabular}
\label{tbl:usedsymbols}
\end{table}

\section{System Model}
\label{sec:system_model}

Fig. \ref{fig:scenario} illustrates the {\em data-intensive \ac{agi} service} model and the {\em cache-enabled MEC network} supporting its delivery, described as follows.

\subsection{Cache-Enabled MEC Network Model} 

The considered cache-enabled \ac{mec} network is modeled by a directed graph $\Set{G} = (\Set{V}, \Set{E})$, with $\Set{V}$ and $\Set{E}$ denoting the node and edge sets, respectively.
Each node $i\in \Set{V}$ represents an edge server, equipped with computation resources (i.e., computing devices) for service function processing.
Each edge $(i, j) \in \Set{E}$ represents a network link supporting data transmission from node $i$ to $j$.
Denote by $\delta^-(i)$ and $\delta^+(i)$ the incoming and outgoing neighbor sets of node $i$, respectively.

Time is slotted, and the available processing and transmission resources are quantified as follows:
\begin{itemize}
    \item Processing capacity $C_i$: the maximum number of instructions (e.g., floating point operations) that can be executed in one time slot at node $i$.
    \item Transmission capacity $C_{ij}$: the maximum number of data units (e.g., packets) that can be transmitted in one time slot over link $(i, j)$.
\end{itemize}

The network nodes are also equipped with storage resources%
\footnote{
    In this paper, ``storage resources'' refers to spaces used for database caching, and data packets associated with service requests are collected in separate spaces, referred to as ``actual queues'' (see Section \ref{sec:queuing}).
}
to cache databases, which are composed of digital objects whose access may be required for service function processing, and we denote by $\Set{K}$ the set of all databases. Define the {\bf caching vector} as
\begin{align} \label{eq:caching_variable}
    \V{x} = \{ x_{i, k} \in \{0, 1\}: i \in \Set{V}, k \in \Set{K} \} 
\end{align}
where $x_{i, k}$ is a binary variable indicating if database (with index $k$) is cached at node $i$ ($x_{i, k} = 1$) or not ($x_{i, k} = 0$). We refer to a node $i$ as a {\em static source} of database $k$ if $x_{i, k} = 1$, and denote by $\Set{V}(k) = \{ i\in \Set{V} : x_{i, k} = 1 \} \subset \Set{V}$ the set of all static sources. A caching vector must satisfy the following storage resource constraint: for $\forall\, i\in \Set{V}$,
\begin{align} \label{eq:storage}
    \sum_{k \in \Set{K}} F_k \, x_{i, k} \leq S_i
\end{align}
where $F_k$ denotes the size of database $k \in \Set{K}$, and $S_i$ the storage capacity of node $i \in \Set{V}$, i.e., the maximum number of static object units (e.g., databases) that can be cached at node $i \in \Set{V}$, respectively. Denote by $\Set{X}$ the set of all caching vectors $\V{x}$ satisfying \eqref{eq:storage}.

There is a cloud data-center in the network, serving as {\em external trusted source} with all databases stored, from which the edge servers can download some databases for caching. We assume that the downloads are supported by another communication system and do not consume the communication resources introduced above.%
\footnote{\label{ft:low_rate}
    Database replacement can be completed with backhaul links between the cloud data-center and edge servers, subject to {\em restricted} transmission rates.
}

\subsection{Data-Intensive Service Model}
\label{sec:service_dag}

\begin{figure}
    \centering
    \includegraphics[width = .95 \columnwidth]{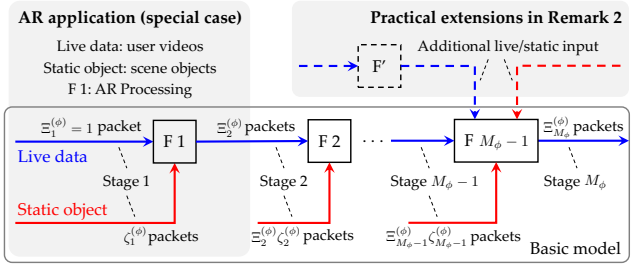}
    \caption{
        The basic model for data-intensive services, including multiple functions (denoted by ``F''), each requiring one live and static streams as inputs. We also depict the AR application as a special case requiring one-step processing, and some practical extensions that can be included in the general DAG model (using F $M_{\phi} - 1$ as an example).
    }
    \label{fig:service}
\end{figure}

Next, we introduce the model for data-intensive service \cite{poularakis2020approximation}, which can be described by a directed acyclic graph (DAG). For ease of exposition, we illustrate the proposed design using a basic model, which simplifies the general model while maintaining the key elements, as illustrated in Fig. \ref{fig:service}. Extensions of the model are described in Remark \ref{rmk:service_DAG}.

Each data-intensive service $\phi$ includes a sequence of $M_{\phi} - 1$ functions, through which the user-specific packets, referred to as {\em live data}, must be processed to produce consumable results, resulting in the end-to-end data stream divided into $M_{\phi}$ stages, and we refer to the input to function $m \in \{1, \cdots, M_{\phi} - 1\}$ as {\em stage $m$ live packet}. In order to process each live packet, the service function requires the other input of a digital object, referred to as {\em static object} \cite{poularakis2020approximation,yeh2021deco}, which can be requested repeatedly and is pre-stored in the network, and we abuse the term {\em stage $m$ static object} to denote the static object input to function $m$ without ambiguity. Each processing step can take place at different network locations hosting the required service functions: for example, in Fig. \ref{fig:scenario}, the two service functions F1 and F2 (in blue) are executed at different edge servers.

Each service function, say the $m$th function of service $\phi$, is specified by $4$ parameters $(\xi^{(\phi)}_m, r^{(\phi)}_m, k^{(\phi)}_m, \zeta^{(\phi)}_m)$, defined as follows (and illustrated in Fig. \ref{fig:service}):
\begin{itemize}
    \item {\bf Object name} $k^{(\phi)}_m$: the name (or index) 
    of the database to which the static object belongs.
    \item {\bf Merging ratio} $\zeta^{(\phi)}_m$: the number of static objects per input live packet.
    \item {\bf Workload} $r^{(\phi)}_m$: the amount of computation resource to process one input live packet.
    \item {\bf Scaling factor} $\xi^{(\phi)}_m$: the number of output packets per input live packet.
\end{itemize}
Note that the ``input live packet'' mentioned in the above definitions refers to the stage $m$ live packet, and we define the {\bf cumulative scaling factor} $\Xi^{(\phi)}_m$ as the number of stage $m$ live packets per stage $1$ live packet (i.e., the initial packet arriving to the network), given by
\begin{align}
    \Xi^{(\phi)}_m = \begin{cases}
    1 & m = 1 \\
    \xi^{(\phi)}_{m-1} \Xi^{(\phi)}_{m-1} & m = 2, \cdots, M_{\phi}
    \end{cases}.
\end{align}

\begin{remark}[Static Object]
\label{rmk:static_object}
Each database can include multiple static objects, e.g., the scene object library in the AR application, and distinct static objects can be required in different service requests. We assume that the live packet and static object belonging to a given request get {\em associated}, i.e., static objects will only be used for the processing of corresponding live packets.
\end{remark}

\begin{remark}[Extended Models]
\label{rmk:service_DAG}
In general, a data-intensive service can be described by a DAG, and we present two practical extensions to the basic model in the following.
(i) For functions with multiple static inputs, we can extend $k^{(\phi)}_m$ and $\zeta^{(\phi)}_m$ (from scalars) to sets.
(ii) For functions with multiple live inputs, we can adopt a tree definition that creates a node for each service function and describes the inputs to it as its child-nodes until reaching the initial/unprocessed live data (i.e., leaf nodes).
The extensions are straightforward and thus omitted.
\end{remark}

\subsection{Client Model}

We define each client $c$ by a $3$-tuple $(s, d, \phi)$, denoting the source node $s$ (where the live packets arrive to the network), the destination node $d$ (where the final packets are requested for consumption), and the requested service $\phi$ (which defines the sequence of service functions and the static objects that are required to process the live packets and create the final packets), respectively.%
\footnote{
    A packet defines the service data unit that can be processed independently, such as a video segment/frame in practice.
}

We refer to a live packet and all static objects required for its processing as belonging to the same {\it packet-level request}.

\subsubsection{Live Packet Arrival}
\label{sec:arrival}

For each client $c$, let $a^{(c)}(t)$ be the number of
live packets arriving at the source node $s$ at time $t$. We assume that the arrival process $\{a^{(c)}(t): t\geq 0\}$ is i.i.d. over time, with a mean arrival rate of $\lambda^{(c)}$ and a maximum number of packet arrivals $A_{\max}^{(c)}$.
Each live packet is immediately admitted to the network upon arrival.

\begin{remark}
In Section \ref{sec:placement} (and the subsequent sections), we assume that there exists a popularity distribution \eqref{eq:popularity_rate} governing the arrival rates of all clients, i.e., $\big\{ \lambda^{(c)}: \forall\, c \big\}$, when designing database placement policies targeting throughput maximization.
\end{remark}

\begin{remark}
\label{rmk:markov_arrival}
We assume i.i.d. arrivals for ease of exposition. Some results, i.e., Theorem \ref{thm:nsr}, \ref{thm:DI-DCNC}, and Proposition \ref{prop:placement}, are valid under the general assumption of Markov-modulated arrivals, i.e., the arrival rate is time-varying and follows a Markov process (see \cite[Section 4.9]{Nee:B10}).
\end{remark}

\subsubsection{Static Object Provisioning}

Upon receiving a packet-level request, for each required static object, one static source is selected to create (by replication) a replica and load it into the network immediately.
Recall that the static object replica gets associated with the corresponding live packet (see also Remark \ref{rmk:static_object}).

\subsection{Queuing System}
\label{sec:queuing}

Each packet (live or static) admitted to the network gets associated with a route encompassing its delivery, and we establish actual queues at each node/link to accommodate packets waiting for processing/transmission at the corresponding locations, respectively.

For each link $(i, j) \in \Set{E}$, we create one {\bf transmission} queue collecting all packets -- regardless of the client, stage, or type (i.e., live or static) -- waiting to cross the link, i.e., packets currently located at node $i$ and having node $j$ as its next hop in the route.
In contrast, two queues are constructed at each node $i \in \Set{V}$:
(i) the {\bf processing} queue collecting the {\em paired} live and static packets concurrently present at node $i$, which are ready for processing, and
(ii) the {\bf waiting} queue collecting the {\em unpaired} live or static packets waiting for their in-transit associates, which are not qualified for processing until joining the processing queue upon their associates' arrival.

\begin{figure}[t]
    \centering
    \includegraphics[width = .99 \columnwidth]{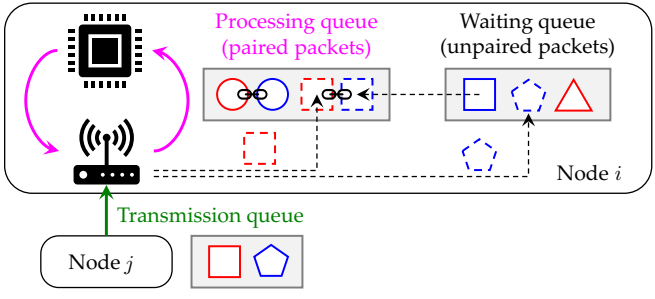}
    \caption{
        Illustration of the queuing system, and in particular, the paired packet queue.
        Different shapes denote packets associated with different requests, blue and red colors the live and static packets, and solid and dashed lines the current and next time slots, respectively.
    }
    \label{fig:queues}
\end{figure}

An illustrative example is shown in Fig. \ref{fig:queues}.
At the current time slot, the paired packet queue (i.e., processing queue) holds a blue and red circle pair, representing live and static packets associated with the same request, which are ready for processing.
At next time slot, when node $i$ receives the red square packet, it gets paired with the blue square packet held in the waiting queue, which together enter the paired packet queue and can be scheduled for processing.

\section{Policy Space and Stability Region}
\label{sec:policy_space}

In this section, we establish an \ac{alg} model for the analysis and optimization of the data-intensive service delivery problem, based on which we characterize the network stability regions.

\subsection{Augmented Layered Graph}

Recent studies have shown that the AgI service (modeled by \ac{sfc}) control problem can be transformed into a packet routing problem on a properly constructed {\em layered graph} \cite{zhang2021multicast}.
While this initial model focuses on the single live service chain routing and processing pipeline, in the data-intensive service involving multiple input streams, a straightforward extension is to incorporate {\em multiple pipelines} into the model.

\subsubsection{Topology of the ALG}

\begin{figure}[t]
    \centering
    \includegraphics[width = .99 \columnwidth]{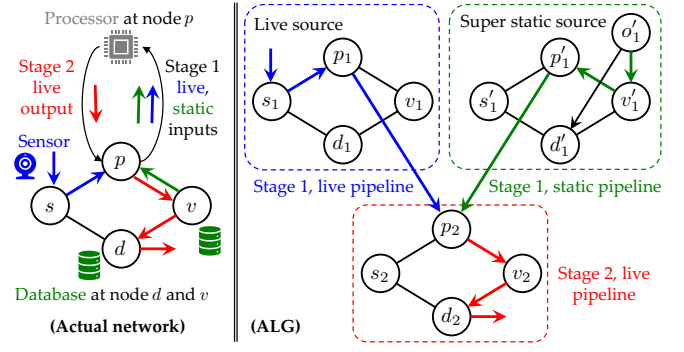}
    \caption{Illustration of the \ac{alg} model.}
    \label{fig:ALG}
\end{figure}

The \ac{alg} associated with service $\phi$ is composed of $M_{\phi}$ layers, indexed by layer $1, \cdots, M_{\phi}$, respectively.
Within each layer, there are two pipelines, termed {\em live} and {\em static} pipelines, respectively, except for the last layer $M_{\phi}$, which only includes the live pipeline.
Each live pipeline has the same topology as the actual network, while each static pipeline (e.g., in layer $m$) includes an additional {\em super static source} node $o_m'$ and the outgoing edges $(o_m', v_m')$ connecting to all static sources $\forall\, v \in \Set{V}(k_m^{(\phi)})$.
We note that:
(i) The live and static pipelines in layer $m$ represent the routing of the stage $m$ live and static packets, respectively.
(ii) With the super static source $o_m'$ in each static pipeline, it is equivalent to assume that {\em $o_m'$ is the only static source of database $k_m^{(\phi)}$}. To wit, if node $o_m'$ can provide static object to node $i_m'$ along the path $(o_m', v_m', \cdots, i_m')$ in the pipeline, we can select the static source $v$ to create the replica and route it to node $i$ along the rest of the path in the actual network, and vice versa.
(iii) The cross-layer edges from layer $m$ to $m+1$ represent processing operations, i.e., the stage $m$ live and static packets pushed through these edges are processed into stage $m+1$ live packets in the actual network.

Fig. \ref{fig:ALG} illustrates the \ac{alg} model using an example of delivering the AR service $\phi$ (as shown in Fig. \ref{fig:service}) over a 4-node network. In the actual network, the stage $m = 1$ live packet, arriving at the source node $s$, and the static object, replicated at the static source $v$, are transmitted to node $p$ (along the blue and green paths), where they also get processed. The produced stage $m+1 = 2$ packet is then transmitted to the destination node $d$ (along the red path).
In the \ac{alg} model: the highlighted links in different pipelines indicate the transmission path of each packet, $(o_1', v_1')$ indicates 
selecting static source $v \in \Set{V}(k_1^{(\phi)}) = \{v, d\}$ to create the replica, and $(p_1, p_2)$ and $(p_1', p_2)$ indicates selecting node $p$ to process the packets.

Formally, given the actual network $\Set{G}$ and the database placement $x$, the ALG of service $\phi$, denoted by $\Set{G}^{(\phi)} = (\Set{V}^{(\phi)}, \Set{E}^{(\phi)})$, is given by
\begin{subequations} \begin{gather}
    \Set{V}^{(\phi)} =  \bigcup_{m = 1}^{M_{\phi}} \Set{V}^{(\phi)}_{\text{L}, m} \cup \bigcup_{m = 1}^{M_{\phi}-1} \Set{V}^{(\phi)}_{\text{S}, m} \\
    \Set{E}^{(\phi)} = \bigcup_{m = 1}^{M_{\phi}} \Set{E}^{(\phi)}_{\text{L}, m} \cup \bigcup_{m = 1}^{M_{\phi}-1} \Set{E}^{(\phi)}_{\text{S}, m} \cup \bigcup_{m = 1}^{M_{\phi}-1} \Set{E}^{(\phi)}_{m, m+1}
\end{gather} \end{subequations}
in which (with L/S in the subscripts denoting live/static)
\begin{align*}
    & \Set{V}^{(\phi)}_{\text{L}, m} = \{ i_m : i \in \Set{V} \},\ \Set{V}^{(\phi)}_{\text{S}, m} = \{ i_m' : i \in \Set{V} \} \cup \{ o_m' \} \\
    &  \Set{E}^{(\phi)}_{\text{L}, m} = \{ (i_m, j_m) : (i, j) \in \Set{E} \}\\
    & \Set{E}^{(\phi)}_{\text{S}, m} = \{ (i_m', j_m') : (i, j) \in \Set{E} \} \cup \{ (o_m', i_m'): i \in \Set{V}(k^{(\phi)}_{m}) \} \\
    & \Set{E}^{(\phi)}_{m, m+1} = \{ (i_m, i_{m+1}), (i_m', i_{m+1}) : i \in \Set{V} \}.
\end{align*}

\begin{remark}
When there are multiple live and static inputs, it is flexible to incorporate more pipelines into each layer. In particular, the proposed \ac{alg} model reduces to the layered graph \cite{zhang2021multicast} when the static object is not relevant (in which case all static pipelines are eliminated).
\end{remark}

\begin{remark}
\label{remark:share}
We note that there are multiple edges in the \ac{alg} associated with the same actual network infrastructure. For example, edge $(i_m, j_m)$, $(i_m', j_m')$ for $\forall\, m$ all correspond to link $(i, j)$, and the operation on any edge consumes the link's communication resource.
\end{remark}

\subsubsection{Flow in the ALG}

Denote by $f_{\imath \jmath} \geq 0$ the network flow associated with edge $(\imath, \jmath) \in \Set{E}^{(\phi)}$ in the ALG, defined as the average packet rate traversing the edge (in {\em packets per slot}). In particular:
\begin{itemize}
    \item $f_{i_m j_m}$ and $f_{i_m' j_m'}$ denote the transmission rates of the stage $m$ live and static streams over link $(i, j) \in \Set{E}$.
    \item $f_{o_m' v_m'}$ denotes the local replication rate of the stage $m$ static stream at static source $v \in \Set{V}(k_m^{(\phi)})$.
    \item $f_{i_m i_{m+1}}$ and $f_{i_m' i_{m+1}}$ denote the processing rates of the stage $m$ live and static streams at node $i \in \Set{V}$.
\end{itemize}

The flows must satisfy the following constraints:

(i) {\em Live flow conservation:} for $\forall\,i\in \Set{V}, 1\leq m \leq M_\phi$,
\begin{align}
\label{eq:flow_conservation_1}
\hspace{-6pt} \sum_{j \in \delta^{+}(i)} f_{i_m j_m} + f_{i_m i_{m+1}} = \sum_{j \in \delta^{-}(i)} f_{j_m i_m} + \xi_{m-1}^{(\phi)} f_{i_{m-1} i_m},
\end{align}
i.e., for stage $m$ live stream, the total outgoing flow rate of transmitted and processed packets is equal to the total incoming flow rate of received packets, and those produced by processing.
Note that processing the input stage $m-1$ live stream at a rate of $f_{i_{m-1} i_m}$ by function $m-1$ produces stage $m$ live stream at a rate of $\xi_{m-1}^{(\phi)} f_{i_{m-1} i_m}$, at node $i$.
Define $f_{i_0 i_1} = f_{i_{M_\phi} i_{M_\phi + 1}} = 0$.

(ii) {\em Static flow conservation:} for $\forall\,i\in \Set{V}, 1\leq m \leq M_\phi-1$,
\begin{align} \label{eq:flow_conservation_2}
    \sum_{j\in \delta^+ (i)} f_{i_m' j_m'} + f_{i_m' i_{m+1}} = \sum_{j \in \delta^- (i) } f_{j_m' i_m'} + f_{o_m' i_m'},
\end{align}
i.e., for stage $m$ static stream, the total outgoing rate of transmitted and processed packets is equal to the total incoming rate of received packets, and those produced by replication. Define $f_{o_m' i_m'} = 0$ for $i \notin \Set{V}(k_m^{(\phi)})$, i.e., nodes that are not static sources.

(iii) {\em Data merging:} for $\forall\,i\in \Set{V}, 1\leq m \leq M_\phi-1$,
\begin{align} \label{eq:merging_ratio}
    f_{i_m' i_{m+1}} = \zeta_m^{(\phi)} f_{i_m i_{m+1}},
\end{align}
i.e., the processing rates of static and live streams at each node $i$ are associated by the merging ratio $\zeta_m^{(\phi)}$, since processing each stage $m$ input live packet requires $\zeta_m^{(\phi)}$ stage $m$ static objects.

\subsection{Policy Space}
\label{sec:policies}

We consider a general policy space for data-intensive service delivery under fixed database placement, encompassing joint packet processing, routing, and replication operations.
To be specific, upon receiving a request, the network controller makes joint decisions on
(i) routing paths and processing locations for live data streams, together with (ii) cache selection and distribution paths for associated data objects.
In addition, each node and link need to decide which packets to schedule for processing and transmission at each time slot.

\subsubsection{Decision Variables}

An admissible policy thus consists of two actions.

\underline{\bf Route Selection:}
For each received request, choose a set of edges in the \ac{alg} and the associated flow $f$ satisfying \eqref{eq:flow_conservation_1} -- \eqref{eq:merging_ratio}, and the decisions follow that:
(i) Cache selection, i.e., selecting the static source to create each static object $k_{m}^{(\phi)}$, is indicated by the replication rate $f_{o_m' v_m'}$.
(ii) Transmission path for each packet (live or static) is composed of the edges with non-zero rates in the corresponding pipeline. While each static packet remembers its own path, note that the initial live packet gets associated with a route that {\em aggregates paths for live packets of all stages}. For example, in Fig. \ref{fig:ALG}, the stage $1$ live packet gets associated with the route including both blue and red edges.
(iii) Processing location selection is indicated by the processing rates $f_{i_m i_{m+1}}$ and $f_{i_m' i_{m+1}}$, and the live and static packets are guaranteed to meet at the same node $i$ due to \eqref{eq:merging_ratio}.

\underline{\bf Packet Scheduling:}
At each time slot, each node $i$ and link $(i, j)$ schedule some packets in the local processing and transmission queues for operation, and the incurred resource consumption shall not exceed the corresponding capacities $C_i$ and $C_{ij}$.
We recall that the processing queue only holds {\em paired} live and static packets.

\subsubsection{Efficient Policy Space} \label{sec:efficient_space}

In this section, we define an efficient policy space, which requires each data packet to follow acyclic paths for delivery, without compromising the performance (e.g., throughput, delay, resource consumption).

More concretely, each request gets associated with an {\bf \ac{ER}} $\sigma$ in the \ac{alg}, defined as:
\begin{itemize}
    
    \item $\sigma$ includes a set of processing locations, denoted by $\big\{ \theta^{(m)} \in \Set{V} : 1 \leq m \leq M_\phi-1 \big\}$, corresponding to the following processing edges in the \ac{alg}
    \begin{align*}
        \big\{ \big( \big[ \theta^{(m)} \big]_m, \big[ \theta^{(m)} \big]_{m+1} \big), \big( \big[ \theta^{(m)} \big]_m', \big[ \theta^{(m)} \big]_{m+1} \big) \big\}_{m=1}^{M_{\phi}-1}
    \end{align*}
    where function $m$ is executed at node $\theta^{(m)}$, and we define $\theta^{(0)} = s$ and $\theta^{(M_{\phi})} = d$.
    
    \item $\sigma$ includes acyclic transmission paths for each packet, and the paths for stage $m$ live and static packets are denoted by:
    \begin{align*}
        & \sigma_{1, m} = \big( \big[ \theta^{(m-1)} \big]_m, \, \cdots,\, \big[ \theta^{(m)} \big]_m \big),\ 1 \leq m \leq M_\phi, \\
        & \sigma_{2, m} = \big( o_m',\, \cdots,\, \big[ \theta^{(m)} \big]_m' \big),\ 1 \leq m \leq M_\phi-1,
    \end{align*}
    respectively.
    
\end{itemize}

In the efficient policy space, for each client $c$, the set of all possible \acp{ER}, denoted by $\Set{F}_c(x)$, is {\em finite}, and the route selection decision can be represented by
\begin{align} 
\V{A}(t) = \{ a^{(c, \sigma)}(t): \sigma \in \Set{F}_c(x), c \}
\end{align}
where $a^{(c, \sigma)}(t) \geq 0$ denotes the number of requests raised by client $c$ at time $t$ that get associated with $\sigma$ for delivery, which satisfies
\begin{align} \label{eq:routing}
    \sum_{\sigma\in \Set{F}_c(x)} a^{(c, \sigma)}(t) = a^{(c)}(t),\ \forall\, c.
\end{align}

Note however that $\Set{F}_c(x)$ includes an exponential number of \acp{ER}, i.e., $|\Set{F}_c(x)| = \Omega(|\Set{V}|^{M_{\phi}-1})$.

\subsection{Network Stability Region} \label{sec:nsr}

In this section, we characterize the {\em network stability regions}, which measure the throughput performance of the edge cloud to support data-intensive service delivery.

\begin{definition}
The network stability region is defined as the set of all arrival vectors $\V{\lambda}$, such that there exists an admissible policy to stabilize the actual queues, i.e.,
\begin{align*}
    \lim_{t\to\infty} \frac{1}{t} \Big[ \sum_{i \in \Set{V}} (R_i(t) + R_i'(t)) + \sum_{(i, j) \in \Set{E}} R_{ij}(t) \Big] = 0
\end{align*}
where $R_i(t)$, $R_i'(t)$ and $R_{ij}(t)$ denote the backlogs of the processing, waiting queues of node $i$, and the transmission queue of link $(i, j)$ at time $t$.
\end{definition}

Let $\Lambda(x)$ and $\Lambda$ be the network stability regions under fixed database placement and when allowing dynamic replacement, characterized in the following.

\begin{theorem} \label{thm:nsr}
For any fixed database placement $x \in \Set{X}$, an arrival vector $\V{\lambda}$ is interior to the stability region $\Lambda(x)$ if and only if for each client $ c$, there exist probability values
\begin{align*}
    \mathbb{P}_c(\sigma) : \sum_{\sigma \in \Set{F}_c(x)} \mathbb{P}_c(\sigma) = 1 \text{ and }
    \mathbb{P}_c(\sigma) \geq 0,
\end{align*}
such that for each node $i$ and link $(i, j)$:
\begin{subequations} \label{eq:nsr} \begin{align}
    \sum_{c} \lambda^{(c)} \sum_{\sigma \in \Set{F}_c(x)} \rho_i^{(c, \sigma)} \mathbb{P}_c(\sigma) & \leq C_i \label{eq:p_resource_nsr} \\
    \sum_{c} \lambda^{(c)} \sum_{\sigma \in \Set{F}_c(x)} \rho_{ij}^{(c, \sigma)} \mathbb{P}_c(\sigma) & \leq C_{ij} \label{eq:t_resource_nsr}
\end{align} \end{subequations}
where $\rho_i^{(c, \sigma)}$ and $\rho_{ij}^{(c, \sigma)}$ denote the processing and transmission resource loads imposed on node $i$ and link $(i, j)$ if a request of client $c$ is delivered by \ac{ER} $\sigma$, given by:
\begin{align} \label{eq:load} 
    \rho_i^{(c, \sigma)} & = \sum_{m} w_{i_m i_{m+1}}^{(c)} \bm{1}_{ \{ (i_m, i_{m+1})\in \sigma \} } \\
    \rho_{ij}^{(c, \sigma)} & = \sum_{m} \big[ w_{i_m j_m}^{(c)} \bm{1}_{ \{ (i_m, j_m )\in \sigma \} }
    + w_{i_m' j_m'}^{(c)} \bm{1}_{ \{ (i_m', j_m')\in \sigma \} } \big] \nonumber
\end{align}
in which
\begin{align} \label{eq:edge_load} 
    w_{\imath \jmath}^{(c)} & = \begin{cases}
    \Xi^{(\phi)}_{m} r^{(\phi)}_{m} & (\imath, \jmath) = (i_m, i_{m+1}) \\
    0 & (\imath, \jmath) = (i_m', i_{m+1}) \\
    \Xi^{(\phi)}_{m} & (\imath, \jmath) = (i_m, j_m) \\
    \Xi^{(\phi)}_{m} \zeta^{(\phi)}_{m} & (\imath, \jmath) = (i_m', j_m')
    \end{cases}.
\end{align}
\end{theorem}

\begin{IEEEproof}
The proof for necessity is given in Appendix \ref{apdx:nsr_path}, and we show the sufficiency by designing an admissible policy DI-DCNC in the subsequent section, and proving that it can support any arrival vector $\V{\lambda} \in \Lambda(x)$.
\end{IEEEproof}

In Theorem \ref{thm:nsr}:
(i) The sum operation in \eqref{eq:load} results from {\em multiple} \ac{alg} edges sharing a {\em common} node/link (see also Remark \ref{remark:share}).
(ii) The probability values $\mathbb{P}_c(\sigma)$ define a randomized policy for route selection, operating as follows: at each time slot, select the \ac{ER} $\sigma \in \Set{F}_{c}(x)$ to deliver the requests of client $c$ with probability (w.p.) $\mathbb{P}_c(\sigma)$.
(iii) The result remains valid under the general assumption of Markov-modulated arrivals, in which case the stability region is defined with respect to time average arrival rate $\lambda^{(c)} = \lim_{T\to\infty} (1/T) \sum_{t=0}^{T-1} a^{(c)}(t)$.

\begin{proposition}
\label{prop:nsr_replacement}
When allowing database replacement, an arrival vector $\V{\lambda}$ is interior to the stability region $\Lambda$ if and only if there exist probability values
\begin{align*}
    \mathbb{P}(x) : \sum_{x \in \Set{X}} \mathbb{P}(x) = 1 \text{ and }
    \mathbb{P}(x) \geq 0,
\end{align*}
and for each database placement $x \in \Set{X}$ and client $c$,
\begin{align*}
    \mathbb{P}_{c, x}(\sigma) : \sum_{\sigma \in \Set{F}_c(x)} \mathbb{P}_{c, x}(\sigma) = 1 \text{ and }
    \mathbb{P}_{c, x}(\sigma) \geq 0
\end{align*}
such that for each node $i$ and link $(i, j)$:
\begin{subequations} \label{eq:nsr_dynamic} \begin{align}
    \sum_{x\in \Set{X}} \mathbb{P}(x) \sum_{c} \lambda^{(c)} \sum_{\sigma\in \Set{F}_{c}(x)} \rho_{i}^{(c, \sigma)} \mathbb{P}_{c, x}(\sigma) & \leq C_{i} \\
    \sum_{x\in \Set{X}} \mathbb{P}(x) \sum_{c} \lambda^{(c)} \sum_{\sigma\in \Set{F}_{c}(x)} \rho_{ij}^{(c, \sigma)} \mathbb{P}_{c, x}(\sigma) & \leq C_{ij}
\end{align} \end{subequations}
with $\rho_i^{(c, \sigma)}$ and $\rho_{ij}^{(c, \sigma)}$ given by \eqref{eq:load}.
\end{proposition}

\begin{IEEEproof}
See Appendix \ref{apdx:nsr_replacement}.
\end{IEEEproof}

In the above proposition, $\mathbb{P}(x)$ denotes the distribution of caching vector $x$ over time, which results from the employed replacement policy; $\mathbb{P}_{c, x}(\sigma)$ plays an equivalent role as $\mathbb{P}_c(\sigma)$ in Theorem \ref{thm:nsr}, i.e., the probability values specifying the route selection policy {\em under placement $x$}.

Comparing Theorem \ref{thm:nsr} and Proposition \ref{prop:nsr_replacement}, we find that allowing database replacement is promising to enlarge the stability region. To wit, setting $\mathbb{P}(x) = \V{1}_{\{ x = x_0 \} }$ in Proposition \ref{prop:nsr_replacement} leads to the same characterization as Theorem \ref{thm:nsr} with $x = x_0$. The result is intuitive assuming that ``each node can complete {\em arbitrary} database replacement {\em immediately}'', since database placement $x$ can be adjusted at each time slot based on the received requests, to optimize the service delivery performance. In Proposition \ref{prop:low_rate}, we will show that the result remains valid under restricted transmission rate for database replacement, in line with footnote \ref{ft:low_rate}.

\begin{remark}
We note that Proposition \ref{prop:nsr_replacement} is derived under the admission control policy of ``selecting the \ac{ER} for each request upon arrival'' (see Section \ref{sec:arrival}), an assumption not necessarily optimal for throughput performance when allowing database replacement, because one can choose to serve the requests when observing a preferable database placement (not upon arrival).
\end{remark}

\section{Multi-Pipeline Flow Control}
\label{sec:DI_DCNC}

In this section, we present the proposed algorithm, referred to as {\em data-intensive dynamic cloud network control} (DI-DCNC), which coordinates routing decisions (including processing decisions by ALG) for the live and static pipelines, as well as traffic scheduling, under fixed database placement.

We first introduce a single-hop virtual system (in Section \ref{sec:virtual_system}) to derive packet routing decisions (in Section \ref{sec:virtual}). Then we present the packet scheduling policy, and summarize the actions taken in the actual network (in Section \ref{sec:ENTO}).

\subsection{Virtual System}
\label{sec:virtual_system}

\subsubsection{Precedence Constraint}

In line with \cite{SinMod:J18,zhang2021multicast}, we establish a virtual network, where the {\em precedence constraint} (that imposes a packet to be transmitted hop-by-hop along its route) is relaxed by allowing a packet upon 
route selection to be immediately inserted into the virtual queues associated with all links in the route. The virtual queue measures the processing/transmission resource load at each node/link in the virtual system, which is interpreted as the {\bf anticipated} resource load at the corresponding network location in the actual network.

For example, suppose that a packet gets associated with the route $(i_1, i_2, i_3)$, then it immediately impacts the queuing states of link $(i_1, i_2)$ and $(i_2, i_3)$ in the virtual system, as opposed to the actual network, where it cannot enter the queue for link $(i_2, i_3)$ before crossing $(i_1, i_2)$.

We emphasize that the virtual system is only used for route selection and not relevant to packet scheduling.

\subsubsection{Virtual Queues}

Denote by $\tilde{Q}_i(t)$ and $\tilde{Q}_{ij}(t)$ the virtual queues for node $i \in \Set{V}$ and link $(i, j) \in \Set{E}$.

The queuing dynamics are given by:
\begin{subequations} \label{eq:virtual_q}
\begin{align}
\tilde{Q}_i(t+1) & = \big[ \tilde{Q}_i(t) - C_i + \tilde{a}_i(t) \big]^+ \\
\tilde{Q}_{ij}(t+1) & = \big[ \tilde{Q}_{ij}(t) - C_{ij} + \tilde{a}_{ij}(t) \big]^+
\end{align}
\end{subequations}
where $C_i$ and $C_{ij}$ are interpreted as the amount of processing/transmission resources that are ``served'' at time $t$;
$\tilde{a}_i(t)$ and $\tilde{a}_{ij}(t)$ are ``additional'' resource loads imposed on the node/link by newly arriving packets.
Recall that each request gets associated with a route for delivery upon arrival, which immediately impacts the queuing states of all links in the route, and thus:
\begin{subequations}
\begin{align}
\tilde{a}_i(t) & = \sum_c \sum_{\sigma \in \Set{F}_c(x)} \rho_i^{(c, \sigma)} a^{(c, \sigma)}(t) \\
\tilde{a}_{ij}(t) & = \sum_c \sum_{\sigma \in \Set{F}_c(x)} \rho_{ij}^{(c, \sigma)} a^{(c, \sigma)}(t)
\end{align}
\end{subequations}
with $\rho_i^{(c, \sigma)}$ and $\rho_{ij}^{(c, \sigma)}$ given by \eqref{eq:load}.

\subsection{Optimal Virtual Network Decisions} \label{sec:virtual}

\subsubsection{Lyapunov Drift Control}

Next, we leverage Lyapunov drift control theory to derive a policy that stabilizes the {\em normalized} virtual queues:
\begin{align*}
\V{Q}(t) & = \Big\{ Q_i(t) \triangleq \tilde{Q}_i(t) \big/ C_i : i\in \Set{V} \Big\} \\
& \hspace{50pt} \cup \Big\{ Q_{ij}(t) \triangleq \tilde{Q}_{ij}(t) \big/ C_{ij} : (i, j) \in \Set{E} \Big\},
\end{align*}
which have equivalent stability properties as the virtual queues (due to the linear scaling) and can be interpreted as {\em queuing delays} in the virtual system.

Define the Lyapunov function as $L(t) \triangleq \| \V{Q}(t) \|^2 / 2$, and the Lyapunov drift $\Delta(\V{Q}(t)) \triangleq L(t+1) - L(t)$.
After some standard manipulations, we derive (in Appendix \ref{apdx:virtual_stability}) the following upper bound of $\Delta(\V{Q}(t))$:
\begin{align} \begin{split} \label{eq:ldp}
& \Delta(\V{Q}(t)) \\
& \hspace{20pt} \leq B - \| \V{Q}(t) \|_1 + \sum_{c} \sum_{\sigma\in \Set{F}_c(x)} O^{(c, \sigma)}(t)\, a^{(c, \sigma)}(t)
\end{split} \end{align}
where $B$ is a constant, and $O^{(c, \sigma)}(t)$ is referred to as the weight of \ac{ER} $\sigma$, given by:
\begin{subequations} \begin{align}
    O^{(c, \sigma)}(t) & = \sum_{i\in \Set{V}} \frac{ Q_i(t) }{ C_i } \, \rho_i^{(c, \sigma)} + \sum_{(i, j)\in \Set{E}} \frac{ Q_{ij}(t) }{ C_{ij} } \, \rho_{ij}^{(c, \sigma)}  \label{eq:ER_weight} \\
    & \hspace{-32pt} = \sum_{i\in \Set{V}} \sum_{m} \tilde{w}_{i_m i_{m+1}}^{(c)}(t) \bm{1}_{ \{ (i_m, i_{m+1})\in \sigma \} } + \sum_{(i, j)\in \Set{E}} \sum_{m} \Big[ \nonumber \\
    & \hspace{-20pt} \tilde{w}_{i_m j_m}^{(c)}(t) \bm{1}_{ \{ (i_m, j_m )\in \sigma \} } + \tilde{w}_{i_m' j_m'}^{(c)}(t) \bm{1}_{ \{ (i_m', j_m')\in \sigma \} } \Big] \label{eq:ALG_weight}
\end{align} \end{subequations} 
in which we plug in \eqref{eq:load}, and
\begin{align} \label{eq:alg_weight}
\hspace{-5pt}
\tilde{w}_{\imath \jmath}^{(c)}(t) = \begin{cases}
\frac{w_{\imath \jmath}^{(c)} Q_i(t)}{C_i} & (\imath, \jmath) = (i_m, i_{m+1}), (i_m', i_{m+1}) \\
\frac{w_{\imath \jmath}^{(c)} Q_{ij}(t)}{C_{ij}} & (\imath, \jmath)  = (i_m, j_m), (i_m', j_m')
\end{cases}
\end{align}
with $\V{w}$ given by \eqref{eq:edge_load}.

The proposed algorithm is designed to minimize the upper bound \eqref{eq:ldp}, or equivalently,
\begin{align} \label{eq:ldp_opt}
    \min_{A(t)}\ \sum_{c} \sum_{\sigma\in \Set{F}_c(x)} O^{(c, \sigma)}(t)\, a^{(c, \sigma)}(t),\ 
    \st\ \eqref{eq:routing}.
\end{align}

\subsubsection{Route Selection}

Given the linear structure of \eqref{eq:ldp_opt}, the optimal route selection decision is given by:
\begin{align} \label{eq:routing_decision}
    a^{\star\, (c, \sigma)}(t) = a^{(c)}(t) \V{1}_{ \{ \sigma = \sigma^\star \} }
\end{align}
where
\begin{align}
    \sigma^\star = \argmin_{\sigma \in \Set{F}_c(x)} \ O^{(c, \sigma)}(t),
\end{align}
i.e., all requests of client $c$ arriving at time $t$ are delivered by the {\em min-\ac{ER}}, i.e., the \ac{ER} with the minimum weight, and the remaining problem is to find the min-ER among the exponential number of \acp{ER} in $\Set{F}_c(x)$.

To this end, we create a {\em weighted \ac{alg}} where each edge $(\imath, \jmath)$ in the \ac{alg} is assigned the weight $\tilde{w}_{\imath \jmath}^{(c)}(t)$ given by \eqref{eq:alg_weight}, under which the weight of the \ac{ER} $\sigma$ \eqref{eq:ALG_weight} equals to the sum of individual edge weights.
In the rest of this section, we propose a dynamic programming algorithm to find the min-ER based on the weighted \ac{alg}.

Define two variables:
\begin{itemize}
    \item The ER weight matrix $W$ of size $M_{\phi} \times |\Set{V}|$, where $W(m, i)$ is the {\em minimum} weight to deliver a stage $m$ live packet to node $i$, {\em optimized over} all previous packet processing/routing/replication decisions.
    \item The processing location matrix $P$ of size $(M_{\phi} - 1) \times |\Set{V}|$, where $P(m, i)$ is the optimal processing location of function $m$, in order to deliver a stage $m+1$ live packet to node $i$.
\end{itemize}
The ultimate goal is to find $W(M_{\phi}, d)$ and the associated \ac{ER} $\sigma^\star$, and the proposed approach is to derive the entries of matrix $W$ row-by-row, or layer-by-layer in the ALG.
To be specific, suppose row $m$ of matrix $W$, i.e., $\{ W(m, j) : j \in  \Set{V}\}$, is given, then we can derive each element on row $m+1$, e.g., $W(m+1, i)$, in two steps:

\begin{figure}[t]
    \centering
    \includegraphics[width = .99 \columnwidth]{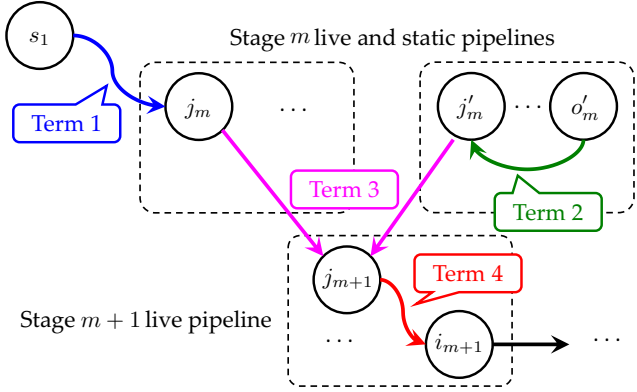}
    \caption{Illustration of the weight components in \eqref{eq:weight_j}.}
    \label{fig:dp}
\end{figure}

\underline{First}, assuming that function $m$ is executed at node $j$, we optimize the cache selection and routing decisions to minimize the weight, i.e.,
\begin{align} \begin{split} \label{eq:weight_j}
W_j (m+1, i) & = W(m, j) + \operatorname{SP}(o_m', j_{m}') \\
& \hspace{30pt} + \tilde{w}_{j_{m} j_{m+1}}(t) + \operatorname{SP}(j_{m+1}, i_{m+1})
\end{split} \end{align}
where $\mathcal{SP}(\imath, \jmath)$ and $\operatorname{SP}(\imath, \jmath)$ denote the shortest path (SP) from node $\imath$ to $\jmath$ in the \ac{alg} and the resulting weight.
As depicted in Fig. \ref{fig:dp}, the four terms represent: (i) the min-weight to deliver a stage $m$ live packet to node $j$, (ii) the min-weight to route the stage $m$ static object to $j$, (iii) the processing load, and (iv) the min-weight to route the produced stage $m+1$ live packet to node $i$, respectively.

\underline{Second}, optimize the processing location decision to minimize the overall weight, i.e.,
\begin{subequations} \label{eq:dp} \begin{gather}
    W(m+1, i) = \min_{j\in \Set{V}} \, W_j (m+1, i), \label{eq:weight} \\ 
    P(m, i) = \argmin_{j\in \Set{V}} \, W_j (m+1, i). \label{eq:location}
\end{gather} \end{subequations}

Repeating the above procedure, we can derive all entries of $W$ and $P$, and the min-\ac{ER} can be found via the following {\em back-tracing} procedure. Starting from destination $d$, the optimal processing location of function $M_{\phi}-1$, $\theta^{(M_{\phi} - 1)}$, can be found by looking up the entry $(M_{\phi} - 1, d)$ of matrix $P$. The remaining problem is to find the optimal decisions to: {\em deliver a stage $M_{\phi}-1$ live packet to node $\theta^{(M_{\phi}-1)}$}, which has the same structure as the original problem, i.e., {\em deliver a stage $M_{\phi}$ live packet to node $d$}. Therefore, we can repeat the above procedure (but assuming $\theta^{(M_{\phi}-1)}$ as the ``new destination'' and $M_{\phi}-1$ as the ``final stage''), until all the processing locations, as well as the routes are determined, as described in Algorithm \ref{alg:dp_opt_tree} (step $5$ to $7$).

\begin{algorithm}[t]

    \textbf{Input}: $\tilde{\bm{w}}(t)$;
    \textbf{Output}: min-ER $\sigma^\star$, optimal weight $W^\star(x)$.

\begin{algorithmic}[1]
    
    \STATE{
    {\bf Initialization:} $W(1, i) \leftarrow \operatorname{SP}(s_1, i_1)$ for $\forall\, i\in \Set{V}$, $\sigma^\star = \mathcal{SP}( [\theta^{(M_{\phi}-1)}]_{M_{\phi}}, d_{M_{\phi}} )$, $\theta^{(M_{\phi}-1)} = P(M_{\phi}-1, d)$.
    }

    \FOR{
    $m = 1, \cdots, M_{\phi}-1$ and $i \in \Set{V}$
    }

    \STATE{
    Calculate row $(m+1)$ for $W$ and row $m$ for $P$ by \eqref{eq:dp}.
    }
    \ENDFOR

    \FOR{
    $m = M_{\phi}-1, \cdots, 1$
    }
    
    \STATE{
    $\theta^{(m-1)} \leftarrow  P(m-1, \, \theta^{(m)})$ (recall that $\theta^{(0)} = s_1$), and
    \begin{align*}
    \sigma^\star \leftarrow \sigma^\star 
    \cup \, \mathcal{SP}(o_m', [\theta^{(m)}]_m') & \cup
    ([\theta^{(m)}]_m', [\theta^{(m)}]_{m+1}) \\
    \cup \, \mathcal{SP}([\theta^{(m-1)}]_{m}, [\theta^{(m)}]_{m})
    & \cup
    ([\theta^{(m)}]_m, [\theta^{(m)}]_{m+1}).
    \end{align*}
    }
    \ENDFOR

    \STATE{
    Return (i) min-ER $\sigma^\star$ and (ii) $W^\star(x) = W(M_{\phi}, d)$.
    }

\end{algorithmic}
\caption{Dynamic Programming to Find the min-ER}
\label{alg:dp_opt_tree}
\end{algorithm}

\subsection{Optimal Actual Network Decisions} \label{sec:ENTO}

Next, we present control decisions in the actual network.

We adopt the route selection decisions made in the virtual network in Section \ref{sec:virtual}. In addition, we adopt extended nearest-to-origin (ENTO) policy \cite{zhang2021multicast} for packet scheduling:

{\em
At each time slot $t$, for each node/link, give priority to the packets which have crossed the smallest number of edges in the \ac{alg} in the corresponding processing/transmission queue.}

We recall that the processing queue only holds paired live and static packets, and the number of crossed hops of a packet pair is equal to that of the live packet component.
The unpaired packets in the waiting queue are not qualified for processing.

To sum up, the proposed DI-DCNC algorithm is described in Algorithm \ref{alg:DI-DCNC}.

\begin{algorithm}[!h]

\begin{algorithmic}[1]
    
    \FOR{$t \geq 0$}
    
    \STATE{
    For each client $c$, all requests arriving at time $t$ are delivered by the min-ER found by Algorithm \ref{alg:dp_opt_tree}.
    }

    \STATE{
    Each link transmits packets and each node processes packet-pairs according to ENTO.
    }

    \STATE{
    Update the virtual queues by \eqref{eq:virtual_q}.
    }

    \ENDFOR

\end{algorithmic}
\caption{DI-DCNC}
\label{alg:DI-DCNC}
\end{algorithm}

\subsection{Performance Analysis}

\subsubsection{Throughput}

Under any fixed database placement, DI-DCNC is throughput optimal, as described in the following theorem.

\begin{theorem} \label{thm:DI-DCNC}
For any fixed database placement $x \in \Set{X}$ and arrival rate $\V{\lambda}$ interior to the network stability region $\Lambda(x)$, all actual queues are rate-stable under DI-DCNC.
\end{theorem}

\begin{IEEEproof}
See Appendix \ref{apdx:throughput-optimality}.
\end{IEEEproof}

\subsubsection{Complexity}

We can take advantage of the following facts to simplify the calculation of \eqref{eq:weight_j} when implementing Algorithm \ref{alg:dp_opt_tree}:
\begin{align*}
    \operatorname{SP}(\imath, \jmath) = w_{\imath \jmath}^{(c)} \operatorname{SP}_0(i, j),\ 
    \forall\, (\imath, \jmath) = (i_m, j_m),\, (i_m', j_m')
\end{align*}
where $\operatorname{SP}_0(i, j)$ denotes the SP distance in the weighted graph which has the same topology as the actual network, with the weight of each edge $(i, j) \in \Set{E}$ given by $Q_{ij}(t)/C_{ij}$. In addition, we note that
\begin{align} \label{eq:static_transmission_load}
    \operatorname{SP}(o_m', j_m') = (\Xi_m^{(\phi)} \zeta_m^{(\phi)}) \min_{v \in \Set{V}(k_m^{(\phi)})} \ \operatorname{SP}_0(v, j).
\end{align}

Therefore, we can implement Algorithm \ref{alg:dp_opt_tree} as follows, and analyze the complexity of each step:
\begin{itemize}
    \item[(i)] Calculate the pairwise SP distance, i.e., $\{ \operatorname{SP}_0(i, j) : (i, j) \in \Set{V} \times \Set{V} \}$ by Floyd-Warshall \cite[Section 25.2]{cormen2009introduction}, with a complexity of $\mathcal{O}(|\Set{V}|^3)$.
    
    \item[(ii)] In each induction step, derive $\operatorname{SP}(o_m', j_m')$ for each node $j \in \Set{V}$ by \eqref{eq:static_transmission_load}, with a complexity of $\mathcal{O}(|\Set{V}|^2)$; then calculate \eqref{eq:weight_j} for each $(i, j)$ pair, with a complexity of $\mathcal{O}(|\Set{V}|^2)$. The total complexity to calculate the entire matrix is $\mathcal{O}(M_{\phi} |\Set{V}|^2)$.
    
    \item[(iii)] Perform back-tracing, with a complexity of $\mathcal{O}(M_{\phi})$.
\end{itemize}
Therefore, the overall complexity of Algorithm \ref{alg:dp_opt_tree} is $\mathcal{O}( |\Set{V}|^3 +  M_{\phi} |\Set{V}|^2 |\Set{C}| )$, with $|\Set{C}|$ denoting the number of clients (because the requests of each client are delivered along the same ER in each time slot).

\subsubsection{Discussions on Delay Performance}
\label{sec:didcnc_delay}

As observed in the numerical experiments (in Section \ref{sec:experiments}), the designed DI-DCNC algorithm can achieve superior delay performance. However, we note that it is not delay optimal, due to the following reasons:

(i) It focuses on queuing delay and neglects the hop-distance of the selected path, which is the dominant component in low-congestion regimes.
(ii) The actual service delay should be taken as the {\em maximum} over the concurrent live and static pipelines, while the \ac{ER} weight \eqref{eq:ALG_weight} depicts the {\em aggregate} delay (i.e., the sum of them).
(iii) The actual queuing delay is impacted by the packet scheduling policy, which does not equal the queuing delay in the simplified virtual system.
(iv) The live and static packets are scheduled separately, lacking an incentive mechanism that jointly treats packet scheduling in all pipelines.

To address the above issues, a promising approach is to keep track of each packet's lifetime \cite{cai2022delay,cai2021delay} and make lifetime-driven decisions, which is left for future work.

\section{Max-Throughput Database Placement}
\label{sec:placement}

The second part of this paper tackles the {\em joint 3C resource orchestration} problem, with this section focusing on the setting of {\em fixed} database placement (and next section designing {\em dynamic} database replacement policies).

\subsection{Problem Formulation}

We aim to design a database placement policy to maximize the throughput, jointly optimized with flow control (i.e., processing/routing) decisions.

\subsubsection{Variables}

We define two variables, representing the database placement and flow control decisions, respectively:
\begin{itemize}
    
    \item {\em Caching vector} $\V{x} = \{ x_{i, k}: i\in \Set{V}, k \in \Set{K} \}$, where $x_{i, k}$ is a binary variable indicating if database $k$ is cached at node $i$ ($x_{i, k} = 1$) or not ($x_{i, k} = 0$).
    
    \item {\em Flow variables} $\V{f} = \{ f_{\imath \jmath}^{(c)} : (\imath, \jmath) \in \Set{E}^{(\phi)}, c \}$ and $\V{f}' = \{ f_{ij}'^{(k)}: (i, j) \in \Set{E}, k \in \Set{K} \}$, where $f_{\imath \jmath}^{(c)}$ denotes the live flow rate of client $c$ on edge $(\imath, \jmath) \in \Set{E}^{(\phi)}$ in the \ac{alg}, and $f_{ij}'^{(k)}$ the static flow rate of database $k$ static objects on link $(i, j) \in \Set{E}$ in the actual network.
    
\end{itemize}

\subsubsection{Constraints}

We impose two classes of constraints on the variables.

(i) {\bf Capacity} constraints limit the multi-dimensional (processing/transmission/storage) network resource usage: the incurred resource consumption shall not to exceed the corresponding capacities, i.e., the processing rate at each node $i\in \Set{V}$ and the transmission rate over each link $(i, j)\in \Set{E}$ must satisfy
\begin{align} \label{eq:resource_constraints} 
\sum_{c, m} r_{m}^{(c)} f_{i_m i_{m+1}}^{(c)} \leq C_i,\ 
\sum_{c, m} f_{i_m j_m}^{(c)} + \sum_{k \in \Set{K}} f_{ij}'^{(k)} \leq C_{ij},
\end{align}
as well as the storage constraint \eqref{eq:storage} at each node $i\in \Set{V}$:
\begin{align*}
\sum_{k \in \Set{K}} F_k \, x_{i, k} & \leq S_i.
\end{align*}

(ii) {\bf Service chaining} constraints impose the relationship between input and output flows as they traverse network nodes and undergo service function processing.
For live flow, the conservation law is given by:
\begin{align} \label{eq:live_conservation}
    & \sum_{j \in \delta^{+}(i)} f_{i_m j_m}^{(c)} + f_{i_m i_{m+1}}^{(c)} = \sum_{j \in \delta^{-}(i)} f_{j_m i_m}^{(c)} + \xi_{m-1}^{(\phi)} f_{i_{m-1} i_m}^{(c)} \nonumber \\
    & \hspace{60pt}  + \lambda^{(c)} \, \bm{1}_{ \{ i_m = s_1 \} },\ \forall\, c, m, i:i_m \ne d_{M_\phi},
\end{align}
and for the destination node:
\begin{align} \label{eq:live_conservation_des}
    f_{d_{M_{\phi}} j_{M_{\phi}}}^{(c)} = 0,\ \forall\, j\in \delta^+(d),
\end{align}
which, compared with \eqref{eq:flow_conservation_1}, includes additional terms representing boundary conditions, i.e., $\lambda^{(c)}$ in \eqref{eq:live_conservation}, and \eqref{eq:live_conservation_des}.

For the static flows (of database $k$), the conservation law can be summarized as follows:
\begin{align} \label{eq:static_conservation}
(1-x_{i, k}) \Big( \sum_{ j\in \delta^+ (i) } f_{ij}'^{(k)} + f_i'^{(k)} - \sum_{ j \in \delta^-(i) } f_{ji}'^{(k)} \Big) = 0
\end{align}
with the processing rate at node $i$ given by
\begin{align} \label{eq:static_processed}
f_i'^{(k)} \triangleq \sum_{c, m} \zeta_{m}^{(\phi)} f_{i_m i_{m+1}}^{(c)} {\bf 1}_{ \{ k_{m}^{(\phi)} = k \} }.
\end{align}
The static flow conservation \eqref{eq:static_conservation} can be described as follows.
For each node $i$ that is {\em not} a static source, i.e., $x_{i, k} = 0$, the static flow of database $k$ must satisfy the flow conservation constraint, i.e.,
\begin{align}
\label{eq:static_conservation_nonsource}
    \sum_{ j\in \delta^+ (i) } f_{ij}'^{(k)} + f_i'^{(k)} = \sum_{ j \in \delta^-(i) } f_{ji}'^{(k)},
\end{align}
and \eqref{eq:static_conservation} is true.
For each static source $i$, in which case $x_{i, k} = 1$, \eqref{eq:static_conservation} is true (because $1 - x_{i, k} = 0$).

Note that \eqref{eq:static_conservation_nonsource} does not necessarily hold at the static sources, because they can perform in-network packet replication: an operation known to violate the flow conservation law \cite{cai2022multicast_arxiv}. In addition, \eqref{eq:static_conservation} is not a linear constraint (in the optimization variables), due to the cross terms of $x$ and $f'$.

\subsubsection{Objective}

We assume that the arrival rates of all clients' requests, $\{ \lambda^{(c)}: \forall\, c\}$, are governed by a popularity distribution.
To be specific, the arrival rates are given by
\begin{align} \label{eq:popularity_rate}
    \big\{ \lambda^{(c)} = p^{(c)} \lambda: \sum_c p^{(c)} = 1 \big\},
\end{align}
and we employ $\lambda$ to measure the throughput performance. This objective is employed targeting better fairness performance, compared to another widely used metric of ``sum flow rate'' that favors clients with lighter resource requirement (since a higher throughput can be achieved by serving such clients provided the same network resource).

\begin{remark}
If the actual popularity distribution is unknown, we use a uniform distribution $\{ p^{(c)} = 1 / |\Set{C}|: \forall\, c\}$ by default. Besides, other than representing the popularity distribution, the values of $p^{(c)}$ can be designed for admission control, customer prioritization, etc..
\end{remark}

\subsection{Proposed Design}

To sum up, the problem is formulated as follows:
\begin{subequations} \label{eq:mip_placement} \begin{align}
    & \max \ \lambda \\
    & \st \hspace{8 pt}  \lambda^{(c)} \geq p^{(c)} \lambda,\ \forall\, c \label{eq:mip_c1} \\
    & \hspace{25 pt} \text{Capacity constraints \eqref{eq:resource_constraints}, \eqref{eq:storage}} \label{eq:resource_constraint} \\ 
    & \hspace{25 pt} \text{Chaining constraints \eqref{eq:live_conservation} -- \eqref{eq:static_processed}} \label{eq:chaining_constraint} \\
    & \hspace{25 pt} x \in \{0, 1\}^{|\Set{K}| \times |\Set{V}|} \text{ and } \V{f}, \V{f}' \succeq 0. \label{eq:data_type}
\end{align} \end{subequations}

We note that \eqref{eq:mip_placement} is a MIP problem due to \eqref{eq:static_conservation}. To improve tractability, we propose to replace \eqref{eq:static_conservation} with the following linear constraint:
\begin{align} \label{eq:static_conservation2}
    \sum_{ j\in \delta^+ (i) } f_{ij}'^{(k)} + f_i'^{(k)} - \sum_{ j \in \delta^-(i) } f_{ji}'^{(k)} \leq C^{\max}_{i, k} \, x_{i, k}
\end{align}
in which
\begin{align} \label{eq:C_max}
    C^{\max}_{i, k} = \sum_{j \in \delta^+(i)} C_{ij} + C_i \, \max_{k_{m}^{(\phi)} = k} \big( \zeta_{m}^{(\phi)} \big/ r_{m}^{(\phi)} \big).
\end{align}

We claim that the resulting MILP problem:
\begin{align} \label{eq:milp_placement}
\hspace{-2pt} \max \ \lambda,\ 
\st\ \eqref{eq:mip_c1}, \eqref{eq:resource_constraint}, \eqref{eq:live_conservation}, \eqref{eq:live_conservation_des},  \eqref{eq:static_conservation2}, \eqref{eq:C_max}, \eqref{eq:static_processed}, \eqref{eq:data_type}
\end{align}
has the same optimal solution as \eqref{eq:mip_placement}.

\begin{IEEEproof}
See Appendix \ref{apdx:milp_equivalence}.
\end{IEEEproof}

In general, the MILP problem \eqref{eq:milp_placement} is still NP-hard, which suffers from high complexity to find the exact solution. However, there are many software toolboxes designed to deal with general MILP problems and can find approximate solutions of improving accuracy with time, and we use the widely adopted \texttt{intlinprog} in MATLAB to implement the proposed design. Furthermore, the derived MILP formulation \eqref{eq:milp_placement} can serve as a good starting point for future studies for approximation algorithm design.

\subsection{Performance Analysis}

In the following, we present an equivalent characterization of the network stability region under fixed database placement $x$.

\begin{proposition} \label{prop:placement}
For any fixed database placement $x \in \Set{X}$, an arrival vector $\V{\lambda}$ is interior to the stability region $\Lambda(x)$ if and only if there exist flow variables $f, f' \succeq 0$ satisfying \eqref{eq:resource_constraints} -- \eqref{eq:static_processed}.
\end{proposition}

\begin{IEEEproof}
In Appendix \ref{apdx:nsr_flow}, we show that the sets of $\V{\lambda}$ described in this proposition and Theorem \ref{thm:nsr} are equal, completing the proof.
Furthermore, the result also applies to Markov-modulated arrivals, as Theorem \ref{thm:nsr} does.
\end{IEEEproof}

By Proposition \ref{prop:placement}, we claim that the proposed database placement policy can lead to maximum throughput.

\section{Database Replacement Policies}
\label{sec:replacement}

In this section, we show the benefits of database replacement under restricted transmission rate (see footnote \ref{ft:low_rate}), which is referred to as {\em replacement rate}, and develop replacement policies to deal with time-varying service demand statistics.

\subsection{Existence of Low-Rate Replacement}

In Section \ref{sec:nsr}, we illustrate the benefits of database replacement assuming ``arbitrary'' and ``immediate'' database replacement, which can impose extremely high requirement for transmission resources. In the following proposition, we show that the same throughput performance can be achieved under arbitrarily low transmission rates.
\begin{proposition}
\label{prop:low_rate}
For any arrival vector interior to the stability region $\Lambda$, there exists a replacement policy achieving an $[\mathcal{O}(T), \mathcal{O}(1/T)]$ tradeoff between average virtual queue backlog and transmission rate.
\end{proposition}

\begin{IEEEproof}
See Appendix \ref{apdx:low-rate-replacement}.
\end{IEEEproof}

A key ingredient of the designed reference policy to reduce the transmission resource requirement is the time frame structure. To be specific, we consider a two-timescale system, where processing and transmission decisions are made on a per time slot basis, and database replacement decisions are made on a per time frame basis, with each frame including $T$ consecutive slots. The replacement is launched at the beginning of each frame, and must be completed at the end of the frame. While the policy remains throughput-optimal regardless of $T$, the required replacement rate can be arbitrarily close to zero by pushing $T\to \infty$, with a tradeoff in the queue backlog (and thus the delay performance).

In the rest of the section, we adopt the time frame structure and design two heuristic database replacement policies, based on estimated {\em popularity distribution} and {\em database score}, respectively. We note that the proposed design is flexible with incorporating advanced prediction techniques, which plays the same role as estimation, but can enhance the timeliness of the quantities.%
\footnote{
    We note that an estimation-based policy derives estimates over frame $\tau$, makes and executes the replacement decisions during frame $\tau + 1$, and the new placement takes effect in frame $\tau + 2$.
}

\subsection{Rate-Based Replacement}

The first policy is a natural extension to the max-throughput database placement policy described in Section \ref{sec:placement}. To handle time-varying demand statistics, we calculate the empirical popularity distribution over each frame $\Set{T}$, i.e.,
\begin{align}
\hat{p}^{(c)} = \frac{ \sum_{t\in \Set{T}} a^{(c)}(t) }{ \sum_{c'} \sum_{t\in \Set{T}} a^{(c')}(t) },
\end{align}
based on which an MILP problem \eqref{eq:milp_placement} can be formulated, whose solution instructs the database replacement at all network nodes.

While straightforward, this policy exhibits three limitations. First, it neglects the existing network loads, which can lead to sub-optimal solution. Second, the new database placement is designed independent with the current placement, neglecting the incurred replacement rate. Finally, it requires solving the MILP problem in an online manner, which can reduce the accuracy of the approximate solution.

\subsection{Score-Based Replacement}

The second policy is motivated by the derived ``min-ER'' rule for route selection, giving rise to a metric, referred to as {\em (database) score}, to evaluate the benefit for each node $i$ to cache a given database.

\begin{figure}
    \centering
    \includegraphics[width = .95 \columnwidth]{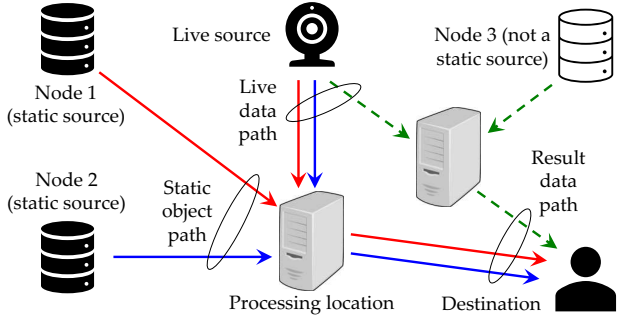}
    \caption{The (i.e., red, blue, and green) min-ERs to deliver the AR service (as shown in Fig. \ref{fig:service}) over the network, assuming that node 1, 2, and 3 are selected as the static source, respectively.}
    \label{fig:score}
\end{figure}

\begin{definition}[Score]
\label{def:file_score}
For each instance, i.e., an arrival packet $\psi$ and given network states (queuing states, database placement), the score of database $k$ at node $i$ is the difference of the min-ER weights assuming node $i$ does not cache the database, and the opposite, given by
\begin{align} \label{eq:score_instant}
    v_{i, k}(\psi) = W^\star( x^-(i, k) ) - W^\star( x^+(i, k) )
\end{align}
where $W^\star(x)$ is the min-\ac{ER} weight given by Algorithm \ref{alg:dp_opt_tree}, with $x^-(i, k)$ and $x^+(i, k)$ denoting the caching vectors equal to $x$, except for $0$ and $1$ in the entry of $x_{i, k}$.
\end{definition}

\begin{figure}
    \centering
    \includegraphics[width = .99 \columnwidth]{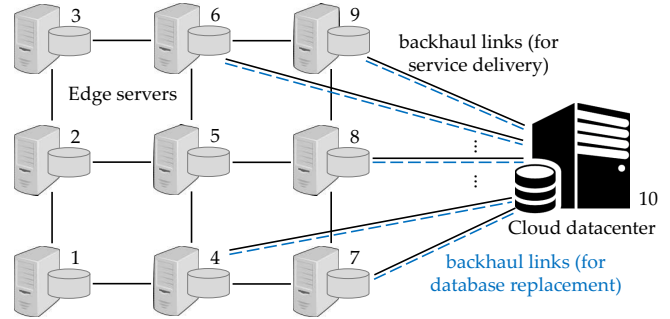}
    \caption{The studied edge cloud, including edge servers (node $1$ to $9$) and a cloud data-center (node $10$). The black and blue edges represent links for user request delivery and database replacement, respectively.}
    \label{fig:network}
\end{figure}

For a given database $k$, at the static source node, the score is the increment of the min-ER weight if it does not cache database $k$; otherwise, the score is the reduction of the min-ER weight if the node caches database $k$.
As illustrated in Fig. \ref{fig:score}: denote by $W_1$, $W_2,$ and $W_3$ the weights of the red, blue, and green ERs, respectively, with $W_3 < W_1 < W_2$, and we note that node $1$ is selected as the static source in the actual network (since node $3$ does not cache the database, and node $2$ incurs a higher weight).
The database score at each node is derived as follows.
Node $1$: if it {\em does not} cache database $k$, node $2$ will be selected, leading to a greater weight of $W_2$, and $v_{1, k} = W_2 - W_1$.
Node $2$: if it {\em does not} cache database $k$, the cache selection decision does not change, leading to the same weight of $W_1$, and $v_{2, k} = W_1 - W_1 = 0$.
Node $3$: if it {\em caches} database $k$, node $3$ will be selected as the static source, leading to a reduced weight of $W_3$, and $v_{3, k} = W_1 - W_3$.

\begin{figure*}[t]
    \centering
    \subfloat[Network stability region.]{
    \includegraphics[width = .9 \columnwidth]{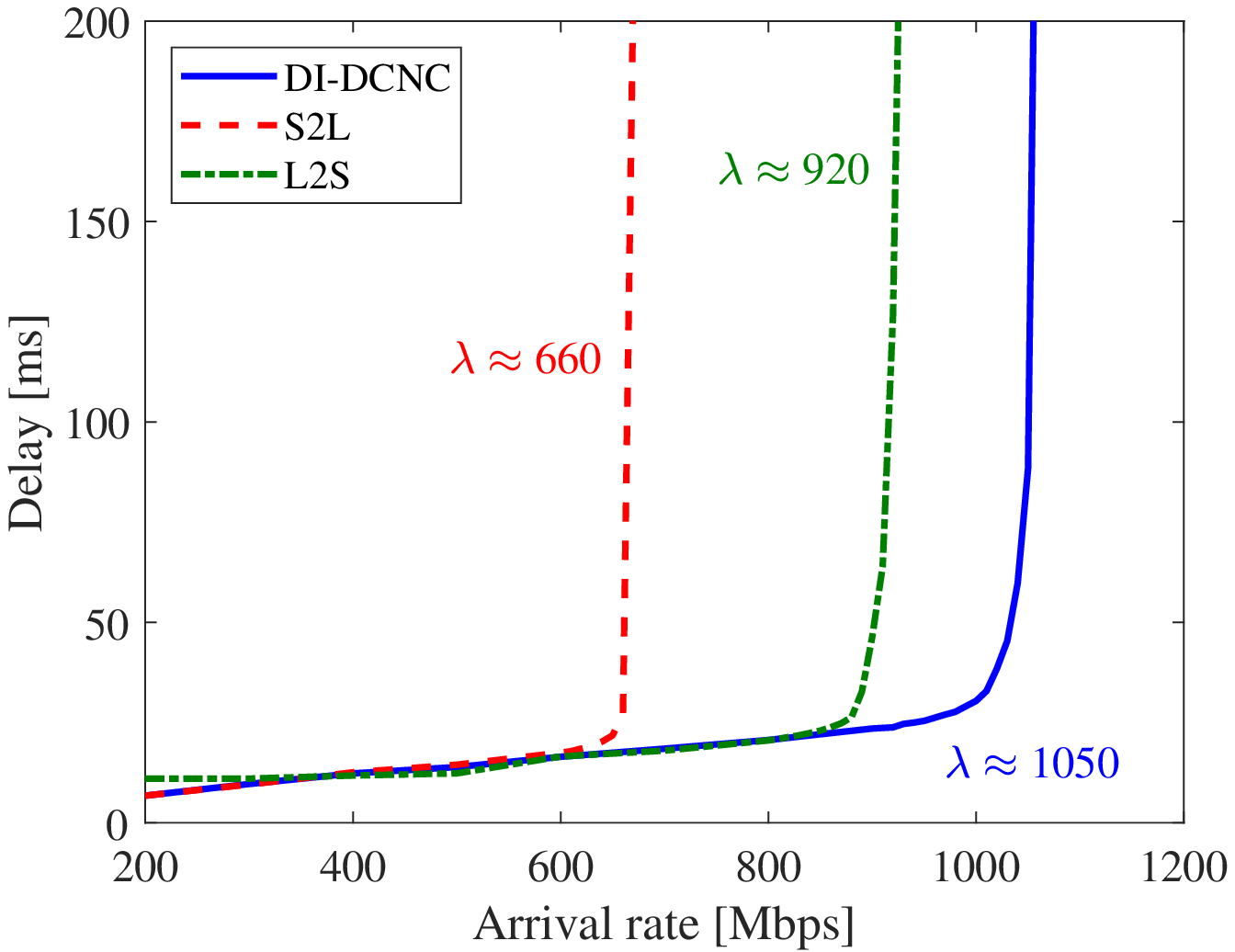}
    \label{fig:network_stability_region}
    }
    \hspace{10pt}
    \subfloat[Resource occupation.]{
    \includegraphics[width = .9 \columnwidth]{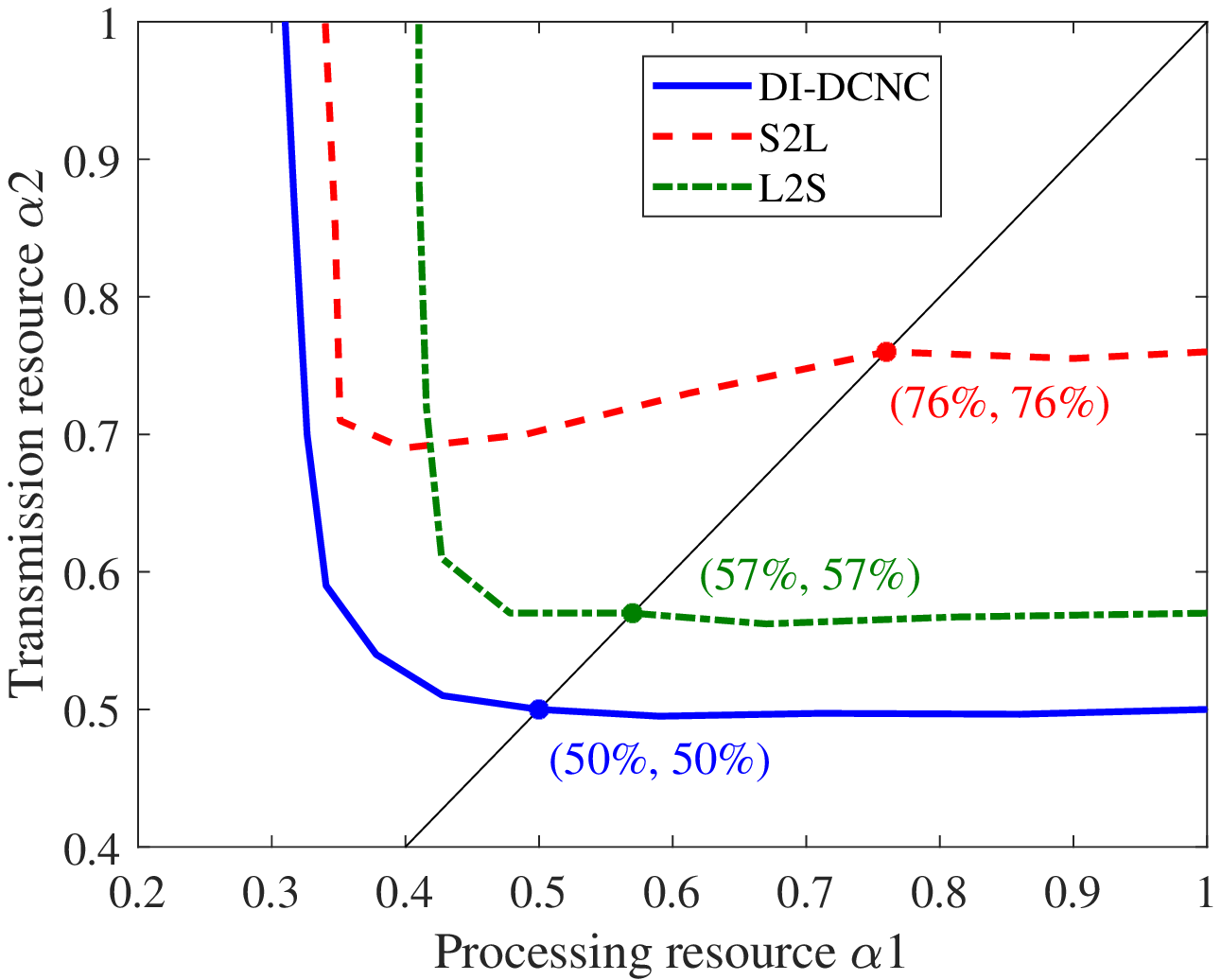}
    \label{fig:tradeoff}
    }
    \caption{Performance of DI-DCNC (under given placement).}
    \label{fig:routing}
\end{figure*}

\begin{table*}[ht]
    \centering
    \caption{Clients, i.e., (source, destination, service), and Service Function Specs, i.e., (scaling factor, workload [\text{GHz}/\text{Gbps}], object name, merging ratio)} \vspace{-5pt}
    \renewcommand\arraystretch{1.5}
    \begin{tabular}{|l|c|c|c|c|} \hline
      Client $(s, d, \phi)$ & $(1, 9, \phi_1)$ & $(3, 7, \phi_2)$ & $(7, 3, \phi_3)$ & $(9, 1, \phi_4)$ \\ \hline
      Func 1 $(\xi_1^{(\phi)}, r_1^{(\phi)}, k_1^{(\phi)}, \zeta_1^{(\phi)})$ & $(0.83,\,7.1,\,1,\,0.92)$ & $(0.94,\, 10.0,\, 3,\, 0.52)$ & $(0.75,\, 8.7,\, 5,\, 1.48)$ & $(0.60,\, 8.4,\, 7,\, 0.91)$ \\ \hline
      Func 2 $(\xi_2^{(\phi)}, r_2^{(\phi)}, k_2^{(\phi)}, \zeta_2^{(\phi)})$ & $(1.06,\, 5.8,\, 2,\, 1.06)$ & $(1.22,\, 7.7,\, 4,\, 0.65)$ & $(1.31,\, 9.2,\, 6,\, 1.97)$ & $(1.34,\, 7.4,\, 8,\,1.22)$ \\ \hline
    \end{tabular} \vspace{-5pt}
    \label{tab:service}
\end{table*}

We note that the definition for score (i) assumes unchanged caching policy at the other network nodes, and (ii) requires finding min-ERs under different placements, which can lead to different processing decisions. In addition, \eqref{eq:score_instant} is the score for a single instance, and the sum score of all instances within a time frame, is a proper metric to evaluate the overall database score, i.e.,\begin{align}
    V_{i, k} = \sum_{t \in \Set{T}} \sum_{\psi \in a(t)} v_{i, k}(\psi).
\end{align}

Given the obtained database scores, we formulate the following optimization problem to decide the database placement at each node $i \in \Set{V}$:
\begin{align} \label{eq:knapsack}
    \max_{x_i \in \{0, 1\}^{|\Set{K}|}} \ \sum_{k \in \Set{K}} V_{i, k}\, x_{i, k},\ 
    \st \ \sum_{k \in \Set{K}} F_k \, x_{i, k} \leq S_i,
\end{align}
i.e., finding the admissible placement to maximize the total score. The above problem, known as {\em 0/1 knapsack problem}, admits dynamic programming solution with pseudo-polynomial complexity $\mathcal{O}( |\Set{K}| S_{i})$ \cite{kleinberg2006algorithm}, and we denote its optimal value and solution by $V_i^\star$ and $x_i^\star$.

Finally, we find the node with the largest total score, i.e., $i^\star = \argmax_i \, V_i^\star$, and {\em only} replace its databases according to $x_{i^\star}^\star$\hspace{1pt}. We refer to the update rule as {\em asynchronous update}, which is in line with the definition of score assuming {\em unchanged} databases at the other network nodes.

There are three factors that can impact the performance of this policy. First, the proposed score metric focuses on each individual node (for tractability), and cannot capture the interaction between network nodes. Second, we use the observed the queuing states to calculate the score, which are also impacted by the database placement. Finally, the asynchronous update can be less efficient for database replacement and lead to slower adaption.

\section{Numerical Results}
\label{sec:experiments}

\subsection{Experiment Setup}

Consider a grid \ac{mec} network composed of $9$ edge servers and a cloud data-center, connected by wired links, as shown in Fig. \ref{fig:network}. Each edge server is equipped with $4$ processors of frequency $2.5$~GHz, and the transmission capacity of each wired link between them is $1$~Gbps. The cloud datacenter is equipped with $8$ identical processors, and the transmission capacity of each backhaul link (for service delivery) is $20$~Mbps. The length of each time slot is $1$~ms.

There are $4$ clients requesting different services, each including $2$ functions, with parameters shown in Table \ref{tab:service}. The size of each packet is $1$~kb, and the arrivals are modeled by i.i.d. Poisson processes with $\lambda$~Mbps.

There are $8$ databases, and each of them has the same size of $F = 1$~Gb, and the storage capacities are in {\em number of databases}. The cloud data-center stores all databases.

\subsection{Multi-Pipeline Flow Control}

We first demonstrate the performance of DI-DCNC under fixed database placement, where database $k = 1, \cdots, 8$ are placed at node $i = 1, \cdots, 4, 6, \cdots, 9$, respectively.

We employ two benchmark algorithms for comparison:\footnote{
    Both S2L and L2S do not consider joint control of multi-pipelines: they focus on the load to transmit either live data, or static objects. 
}
\begin{itemize}
    \item {\em Static-to-live (S2L),} which makes individual routing decisions for live data \cite{zhang2021multicast}, and selects the {\em nearest} static sources to create objects for the processing locations along {\em shortest paths} (in weighted ALG).
    \item {\em Live-to-static (L2S),} which brings live data to the {\em nearest} (in weighted ALG) static source for processing.
\end{itemize}

\subsubsection{Network Stability Region}

First, we study the network stability regions attained by the algorithms, and depict the average delay under different arrival rates (which is set to $\infty$ if the network is not stable).

As shown in Fig. \ref{fig:network_stability_region}, DI-DCNC attains good delay performance over a wide range of arrival rates; when $\lambda$ crosses a {\em critical} point ($\lambda_1 \approx 1.05$ Gbps), at which point the average delay blows up, indicative of the stability region boundary. Similar behaviors are observed from S2L and L2S, and we find that DI-DCNC outperforms them in terms of the achieved throughput: $1.05$ Gbps (DI-DCNC) $>$ $920$ Mbps (L2S) $>$ $660$ Mbps (S2L). Therefore, compared to S2L and L2S, DI-DCNC can better exploit network resources to improve the throughput. We clarify that the performance of S2L improves when increasing the transmission resources, and can outperform L2S under some configurations \cite{cai2022CCC_arXiv2}.

We also notice that the delay attained by DI-DCNC is very similar and not lower than both benchmarks in low-congestion regimes. As discussed in Section \ref{sec:didcnc_delay}, DI-DCNC is designed to reduce the {\em aggregate queuing} delay of both live and static data pipelines; such objective -- while closely related to (especially in high-congestion regimes) -- is not exactly equivalent to the actual service delay, which (i) depends on the maximum delay between the two concurrent pipelines, and (ii) depends also on the hop-distance of the path, neglected by DI-DCNC.

\begin{figure*}[t]
    \centering
    \subfloat[Network stability region (error-bar represents the standard deviation).]{
    \includegraphics[width = .9 \columnwidth]{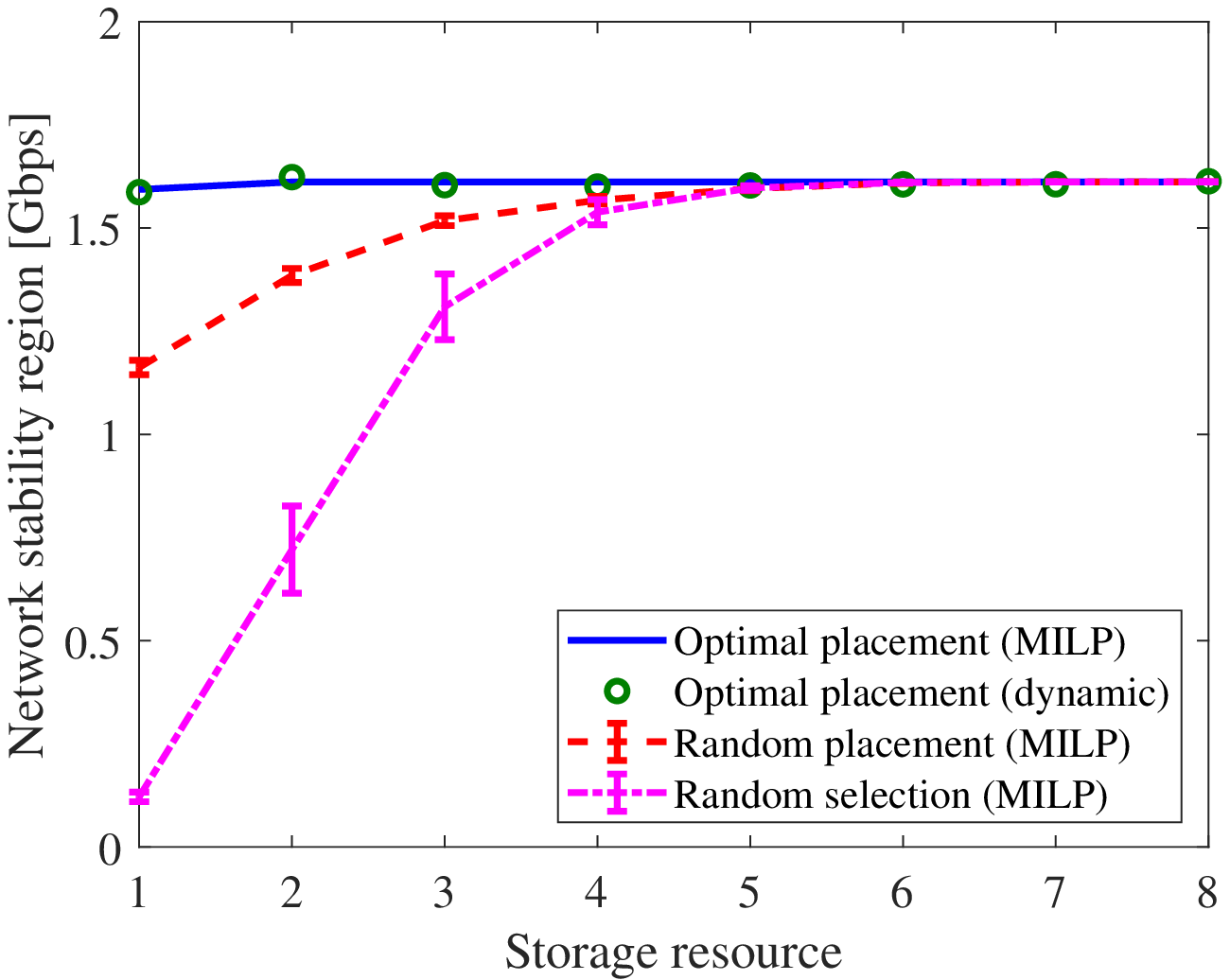}
    \label{fig:placement_stability_region}
    }
    \hspace{10pt}
    \subfloat[Resource occupation.]{
    \includegraphics[width = .9 \columnwidth]{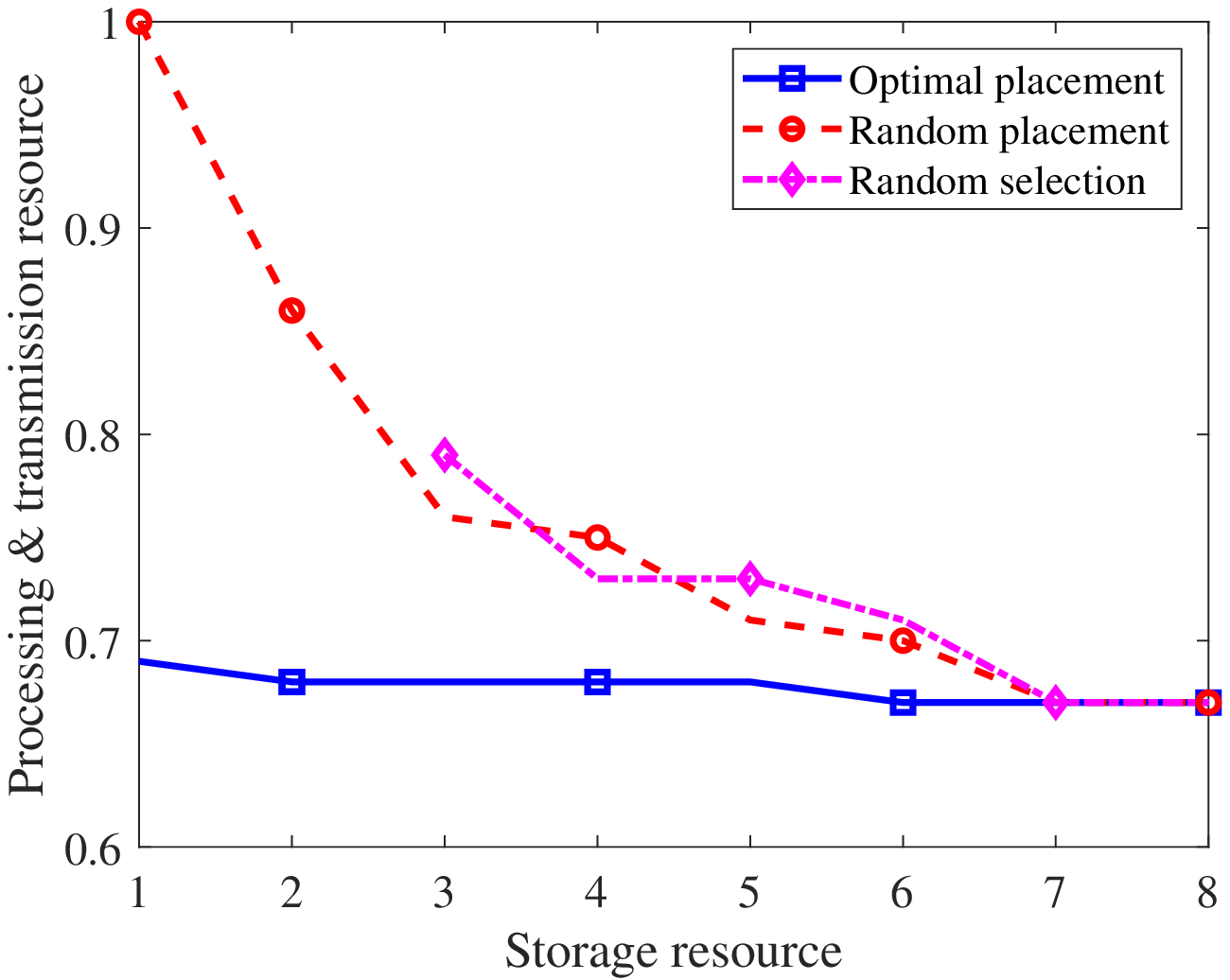}
    \label{fig:placement_resource_occupation}
    }
    \caption{Performance of max-throughput database placement (fixed).}
    \label{fig:placement}
\end{figure*}

\subsubsection{Resource Occupation}

Next, we compare the resource occupation of the algorithms. Assume that the available processing/transmission capacities at each node/link are $\alpha_1$ and $\alpha_2$ (in percentage) of corresponding maximum budgets, respectively, and we define the {\em feasible region} as the collection of $(\alpha_1, \alpha_2)$ pairs under which the delay requirement is fulfilled. Assume $\lambda = 500$~Mbps and average delay requirement $= 30$~ms.

Fig. \ref{fig:tradeoff} depicts the feasible regions attained by the algorithms. Since a lower latency can be attained with more resources, i.e., $(\alpha_1, \alpha_2) \to (1, 1)$, the feasible regions are to the upper-right of the border lines.%
\footnote{
    We note that the feasible region achieved by S2L is not convex.
}
As we can observe, DI-DCNC can save the most network resources. In particular, when $\alpha_1 = \alpha_2 = \alpha$, the resource saving ratios, i.e., $1 - \alpha$, of the algorithms are: $50\%$ (DI-DCNC) $>$ $43\%$ (S2L) $>$ $24\%$ (L2S). Another observation is: S2L is transmission-constrained, compared to its sensitivity to processing capacity. To wit: in the horizontal direction, it can achieve a maximum saving ratio of around $70\%$ (when $\alpha_2 = 1$), which is comparable to DI-DCNC; while the maximum saving is $25\%$ for transmission resources (when $\alpha_1 = 1$), and there is a large gap between the attained performance and that of DI-DCNC ($\approx 50\%$). The reason is that S2L ignores the transmission load of static objects, leading to additional transmission resource consumption. In contrast, L2S is processing-constrained, because only the processing resources at static sources are available for use.

\subsection{Joint 3C Resource Orchestration}

Then, we evaluate the proposed database placement and replacement policies, using DI-DCNC for flow control.

\subsubsection{Fixed Placement}

First, we focus on the setting of fixed database placement. We assume each edge server can cache $S$ databases, and employ two random policies as benchmarks:
\begin{itemize}
    \item {\em Random placement:} each edge server caches $S$ different databases, that are jointly selected to maximize the database diversity in all edge servers.%
    \footnote{
    The cached databases at all edge servers are selected as follows. Generate a random permutation of sequence $\{ 1, \cdots, |\Set{K}| \}$, repeat it for $\lceil |\Set{V}| S / |\Set{K}| \rceil$ times, and denote the resulting sequence by $\Set{D}$ and its $i$th element by $\Set{D}_i$. Then, databases $\Set{D}_{(i-1) S + 1}, \cdots, \Set{D}_{iS}$ are cached at edge server $i$. Note that they are different databases because every sub-sequence in $\Set{D}$ with length $S \leq |\Set{K}|$ does not include repeated value.
    }
    \item {\em Random selection:} each edge server randomly selects $S$ different databases to cache.
\end{itemize}

Similar performance metrics, i.e., network stability region and resource occupation, are used to evaluate the policies, and the results are summarized in Fig. \ref{fig:placement}.

(i) {\em Network Stability Region:}
In Fig. \ref{fig:placement_stability_region}, we present the stability region attained by the proposed placement policy, together with the optimal throughput derived from the MILP \eqref{eq:milp_placement} (for the random policies with given placement $x$, the MILP problem reduces to linear programming).

For each policy, the attained throughput grows with the storage resource, and the proposed policy outperforms the random benchmarks, especially with limited storage resources. For example, when $S = 1$, the proposed policy achieves the best performance ($\approx 1.59$ Gbps), which is $37\%$ better than {\em random placement} and far beyond {\em random selection}. The result validates the effects of ``where to cache the databases'', in addition to ``which databases to cache'', on the throughput performance. Finally, the stability regions collected in the experiments agree with the results derived from MILP, validating Proposition \ref{prop:placement}.

(ii) {\em Resource Occupation:}
Next, we study the tradeoff between multi-dimensional (processing, transmission and storage) network resources, assuming $\lambda = 1$ Gbps, average delay requirement $= 30$ ms, and $\alpha_1 = \alpha_2 = \alpha$. For each random benchmark, we select a representative placement that attains a throughput performance closest to the corresponding mean value (as shown in Fig. \ref{fig:placement_stability_region}).

\begin{figure*}[t]
    \centering
    \subfloat[Network stability regions (frame size $= 100\,\text{s}$).]{
    \includegraphics[width = .9 \columnwidth]{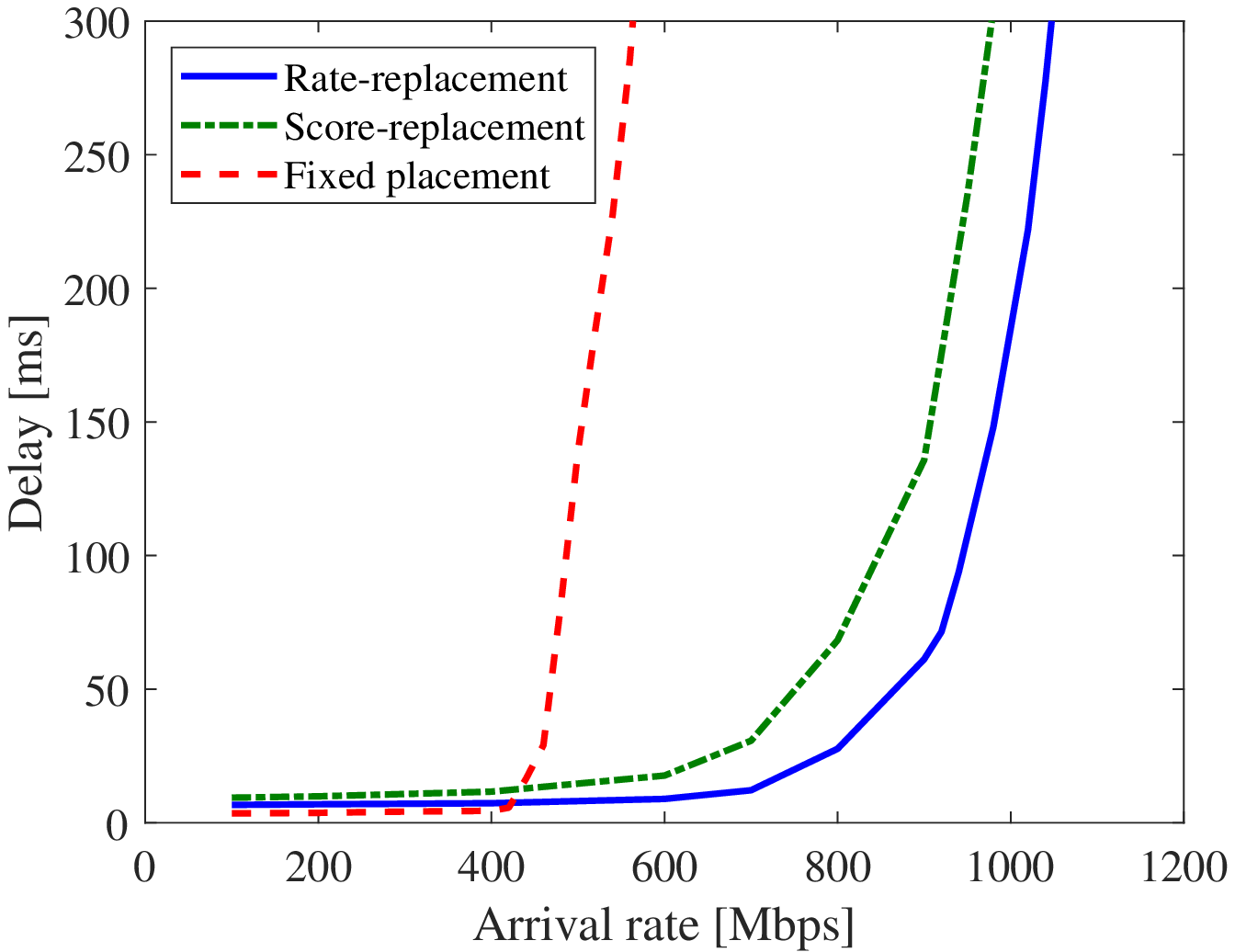}
    \label{fig:delay_replacement}
    }
    \hspace{10pt}
    \subfloat[Effects of frame size on throughput and cost.]{ 
    \includegraphics[width = .9 \columnwidth]{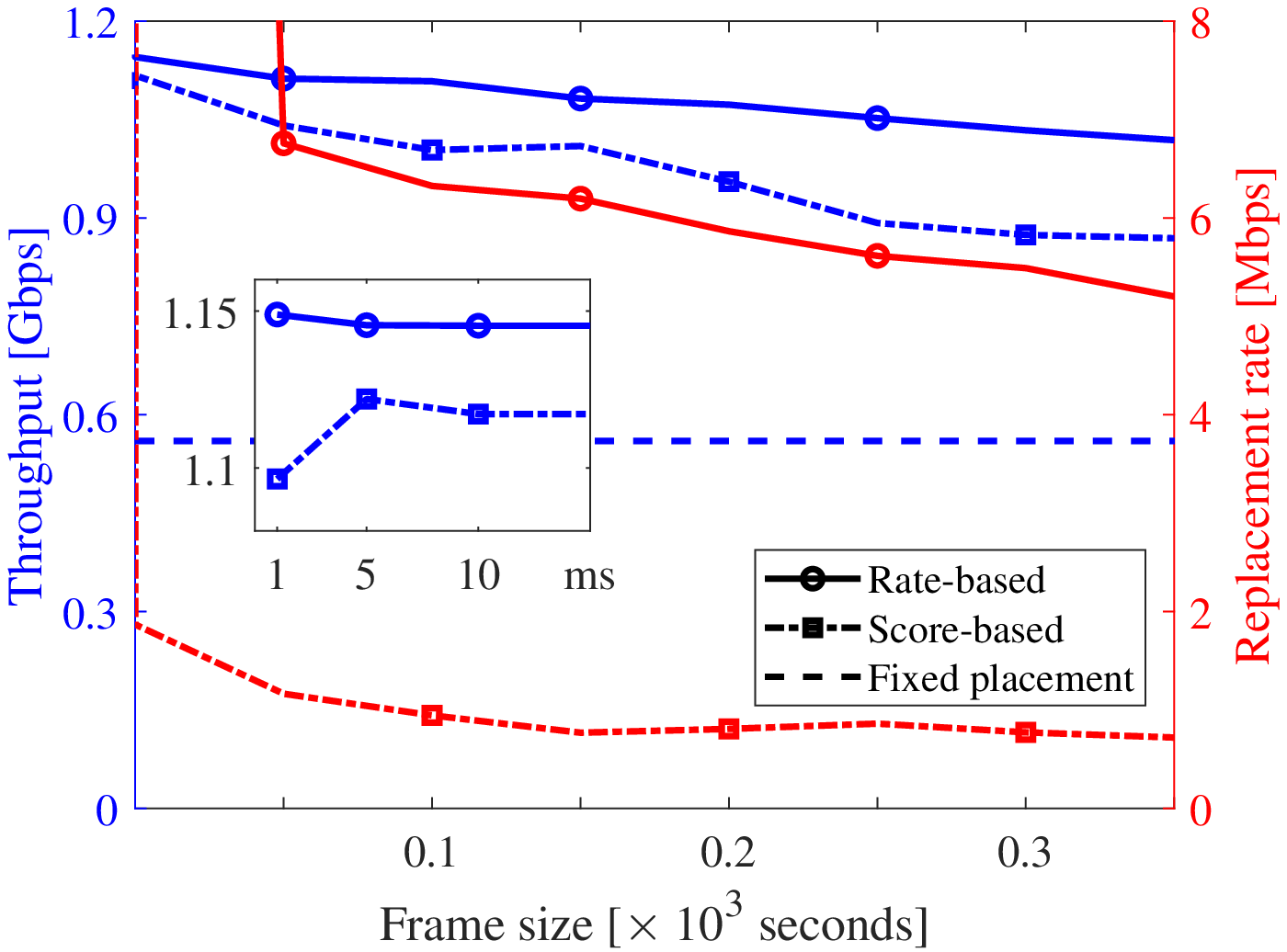}
    \label{fig:window_replacement}
    }
    \caption{Performance of rate- and score-based replacement policies.
    }
    \label{fig:replacement}
\end{figure*}

As depicted in Fig. \ref{fig:placement_resource_occupation}, with more storage resources at the edge servers, a larger saving ratio of processing and transmission resources can be attained. In particular, the performances of the three policies converge when each node has sufficient space to cache all databases (i.e., $S = |\Set{K}|$). When the storage resource is limited (e.g., $S = 1$), the proposed design can achieve a processing/transmission resource saving ratio ($69\%$) close to the optimal value (around $67\%$), outperforming the two random benchmarks.

\subsubsection{Replacement Policies}

Finally, we evaluate the proposed database replacement policies. The arrivals are modeled by Markov-modulated process, i.e., the arrival rate follows a Markov process (described in the following), and the arrivals at each time slot is a Poisson variable. Popularity distribution at each time slot is a {\em permutation} of the Zipf distribution with $\gamma = 1$ \cite{poularakis2020approximation}, with the initial order generated randomly; at each time slot, w.p. $10^{-6}$, we randomly select a number $\varphi \in \{1, 2, 3\}$, and exchanges the popularity ranking of the $\varphi$-th and $(\varphi + 1)$-th most popular services, i.e., their arrival rates.%
\footnote{
    Under this setting, the expected time for the popularity distribution to change is $10^6 \,\text{ms} \approx 15\, \text{min}$; in addition, the time average arrival rate for all services are equal (i.e., uniform popularity).
}

Each edge server, except node $i = 4, 5, 6$, is allowed to cache $S = 1$ database,%
\footnote{
    Under this setting, the total storage resources at the edge servers are $6$ and cannot support caching all the databases (since $|\Set{K}| = 8$), which is a common scenario in practical systems.
}
and the database placement is given by the proposed max-throughput policy assuming uniform popularity distribution.
The two proposed replacement policies are evaluated, and for fair comparisons, we set equal running time for the two policies (to calculate the estimated quantities and solve corresponding problems).

The results are shown in Fig. \ref{fig:replacement}.
Fig. \ref{fig:delay_replacement} plots the throughput performance of the two replacement policies, both of which effectively boost the throughput performance ($\approx 1$ Gbps) compared to fixed placement ($\approx 0.5$ Gbps),
despite the slightly worse delay performance in low-congestion regimes (e.g., $\lambda \leq 400$ Mbps).
Fig. \ref{fig:window_replacement} demonstrates the effects of (time) frame size on the throughput performance and replacement rate requirements of the two policies.
With growing frame size, the frequency of database replacement reduces, leading to sub-optimal throughput, with reduced replacement rate requirement.%
\footnote{
    The selection of frame size controls the tradeoff between the accuracy and timeliness of the estimate. The tradeoffs can be observed from the score-based policy (selecting the frame size as $5$ ms achieves maximum throughput), but is negligible in the rate-based policy.
}
Comparing the two proposed policies, we find that: rate-based policy can achieve better throughput, which is also less sensitive to the frame size (note that the blue-solid curve is more flat); while score-based policy achieves a much lower cost, which updates the placement based on current network states and in an asynchronous fashion.

\section{Conclusions}
\label{sec:conclusions}

We investigated the problem of joint 3C control for the online delivery of data-intensive services, in which each service function can include multiple input streams. 
We first characterized the network stability regions based on the proposed \ac{alg} model, which incorporates multiple pipelines to deal with corresponding inputs.
We then addressed two problems:
(i) multi-pipeline flow control,
in which we derive a throughput-optimal control policy DI-DCNC to coordinate processing, routing and cache selection decisions for multiple pipelines,
and (ii) joint 3C resource orchestration, in which we designed a max-throughput database placement policy that jointly optimizes 3C decisions, as well as the rate-based and score-based policies for database replacement.
Via numerical experiments, we demonstrated the superior performance of multiple pipeline coordination and integrated 3C design in delivering next-generation data-intensive real-time stream-processing services.

\bibliographystyle{IEEEtran}
\bibliography{IEEE_abrv,AgI}

\clearpage

\appendices

\section{Necessity of Route-based Characterization}
\label{apdx:nsr_path}

Consider an arrival process $a(t) = \{ a^{(c)}(t) : \forall\, c \}$ and the policy able to support it, i.e., keeping the actual queues stable. Let $A^{(c, \sigma)}(t)$ be the amount of requests of client $c$ that are successfully delivered via \ac{ER} $\sigma$ by time $t$.

By the conservation of packets:
\begin{align} \label{eq:data_conservation}
    Y^{(c)}(t) + \sum_{\sigma \in \Set{F}_c(x)} A^{(c, \sigma)}(t) = \sum_{\tau = 0}^{t-1} a^{(c)}(\tau)
\end{align}
where $Y^{(c)}(t)$ is the amount of live packets associated with client $c$ and staying in the network by time $t$.
Divide it by $t$, let $t\to\infty$, and note the following facts:
\begin{align}
    \lim_{t\to\infty} \frac{Y^{(c)}(t)}{t} = 0,\ 
    \lim_{t\to\infty} \frac{1}{t} \sum_{\tau = 0}^{t-1} a^{(c)}(\tau) = \lambda^{(c)}
\end{align}
because the queues are stable, and the arrival process is i.i.d. (the relationship is the definition of $\lambda^{(c)}$ if the arrival is modeled by Markov-modulated process, i.e., time average arrival rate).
In addition, define
\begin{align}
    \lambda^{(c, \sigma)} \triangleq \lim_{t\to\infty} \frac{A^{(c, \sigma)}(t)}{t},\ 
    \mathbb{P}_c^{(\sigma)} \triangleq \frac{\lambda^{(c, \sigma)}}{\lambda^{(c)}},
\end{align}
and \eqref{eq:data_conservation} becomes
\begin{align}
    \sum_{\sigma\in \Set{F}_c(x)} \mathbb{P}_c^{(\sigma)} = 1.
\end{align}

On the other hand, the total resource consumption at each interface $e$ (each interface $e$ can be a node $i \in \Set{V}$ or a link $(i, j) \in \Set{E}$, which uses the available processing/transmission resources to handle the packets) cannot exceed the corresponding capacity:
\begin{align}
\sum_{c} \sum_{\sigma \in \Set{F}_c(x)} \rho_e^{(c, \sigma)} A^{(c, \sigma)}(t) \leq C_e\, t.
\end{align}
Divide it by $t$, let $t\to\infty$:
\begin{align}
\sum_{c} \sum_{\sigma \in \Set{F}_c(x)} \rho_e^{(c, \sigma)} \lim_{t\to\infty} \frac{A^{(c, \sigma)}(t)}{t} 
& = \sum_{c} \sum_{\sigma \in \Set{F}_c(x)} \rho_e^{(c, \sigma)} \lambda^{(c, \sigma)} \nonumber \\
& \hspace{-60pt} = \sum_{c} \lambda^{(c)} \sum_{\sigma \in \Set{F}_c(x)} \mathbb{P}_c^{(\sigma)} \rho_e^{(c, \sigma)}
\leq C_e,
\end{align}
which is \eqref{eq:nsr}, concluding the proof.

\section{Throughput-optimality of DI-DCNC}
\label{apdx:throughput-optimality}

We will prove the theorem in two steps, by showing (i) the stability of the virtual queues under the virtual routing decisions \eqref{eq:routing_decision}, and (ii) the stability of the actual queues under DI-DCNC (which incorporates ENTO scheduling).

\subsection{Stability of Virtual Queues}
\label{apdx:virtual_stability}

Suppose the arrival vector $\V{\lambda}$ satisfies \eqref{eq:nsr}, i.e., there exists probability values $\mathbb{P}_c(\sigma)$ such that for each interface $e$:
\begin{align}
\sum_{c} \lambda^{(c)} \sum_{\sigma \in \Set{F}_c(x)} \rho_e^{(c, \sigma)} \mathbb{P}_c(\sigma) & \leq C_e,
\end{align}
or equivalently: there exists $\epsilon \geq 0$, such that
\begin{align}
\label{eq:slackness}
C_e - \sum_{c} \lambda^{(c)} \sum_{\sigma \in \Set{F}_c(x)} \rho_e^{(c, \sigma)} \mathbb{P}_c(\sigma)
& \geq \epsilon C_e
\end{align}
for $\forall\, e$ (note that $\epsilon = 0$ satisfies the condition).

Define a reference policy $*$ operating as follows: at each time slot, select the ER $\sigma$ to deliver the requests of client $c$ w.p. $\mathbb{P}_c^{(\sigma)}$.
Denote the associated route selection decision by $a^{* \, (c, \sigma)}(t)$, and the additional resource load imposed on interface $e$ is given by (in the virtual system):
\begin{align} \begin{split}
\E{ \tilde{a}_e^*(t) } & = \sum_{c} \sum_{\sigma \in \Set{F}_c(x)} \rho_e^{(c, \sigma)} \E{ a^{* \, (c, \sigma)}(t) } \\
& = \sum_{c} \lambda^{(c)} \sum_{\sigma \in \Set{F}_c(x)} \mathbb{P}_c^{(\sigma)} \rho_e^{(c, \sigma)},
\end{split} \end{align}
and thus:
\begin{align}
\frac{ C_e - \E{ \tilde{a}_e^*(t) } }{ C_e } \geq \epsilon.
\end{align}

On the other hand, the virtual queue drift can be derived as follows.
Take square of \eqref{eq:virtual_q}:
\begin{align}
\tilde{Q}_e(t+1)^2 & \leq \left[ \tilde{Q}_e(t) - C_e + \tilde{a}_e(t) \right]^2 \nonumber \\
& = \tilde{Q}_e(t)^2 + \left[ C_e - \tilde{a}_e(t) \right]^2 - 2 \left[ C_e - \tilde{a}_e(t) \right] \tilde{Q}_e(t) \nonumber \\
& \leq \tilde{Q}_e(t)^2 + 2 B_e - 2 \left[ C_e - \tilde{a}_e(t) \right] \tilde{Q}_e(t) \label{eq:quadratic_bound}
\end{align}
with
\begin{align}
B_e = \frac{1}{2} \max \left\{ C_e, \sum_{c} A_{\max}^{(c)} \max_{\sigma \in \Set{F}_c(x)} \rho_e^{(c, \sigma)} \right\}^2.
\end{align}
Therefore,
\begin{align} \begin{split}
\Delta (\tilde{Q}_e(t)) & \triangleq \frac{ \tilde{Q}_e(t+1)^2 - \tilde{Q}_e(t)^2 }{2} \\ 
& \leq B_e - \left[ C_e - \tilde{a}_e(t) \right] \tilde{Q}_e(t),
\end{split} \end{align}
and the overall {\em normalized} drift is bounded by
\begin{subequations} \label{eq:one_slot_drift}
\begin{align}
    \Delta (\V{Q}(t)) & \leq B - \sum_{ e \in \Set{V}\,\cup\,\Set{E} } \frac{ C_e - \tilde{a}_e(t) }{C_e} \, Q_e(t) \\
    & \overset{(\text{a})}{\leq} B - \sum_{ e \in \Set{V}\,\cup\,\Set{E} } \frac{C_e - \tilde{a}_e^*(t)}{C_e} \, Q_e(t)
\end{align}
\end{subequations}
with $B = \sum_{e \in \Set{V}\,\cup\,\Set{E}} B_e / C_e^2$, and inequality (a) is due to the proposed route selection decision {\em minimizing} the bound.

\subsubsection{I.I.D. Arrival}

Take expectation of \eqref{eq:one_slot_drift}, and we obtain:
\begin{align}
\E{ \Delta (\V{Q}(t)) }
& \overset{(\text{b})}{\leq} B - \sum_{ e \in \Set{V}\,\cup\,\Set{E} } \frac{C_e - \E{\tilde{a}_e^*(t)}}{C_e} \, \E{Q_e(t)} \nonumber \\
& \leq B - \epsilon \, \| \E{\V{Q}(t)} \|_1.  \label{eq:one_slot_drift_2}
\end{align}
In inequality (b), we use the fact that the decisions made by $*$ are independent with the current queuing status, and thus expectation multiplies. In fact, \eqref{eq:one_slot_drift_2} implies that the virtual queues are rate-stable \cite[Section 3.1.4]{Nee:B10}.

\subsubsection{Markov-Modulated Arrival}
\label{apdx:Markov_arrival}

Under Markov-modulated arrival process, \eqref{eq:one_slot_drift_2} is violated because $\tilde{a}_e^*(t)$ and $Q_e(t)$ are dependent (since both of them depend on the arrival in previous time slot, i.e., $\tilde{a}_e^*(t-1)$), and we employ the {\em multi-slot drift} technique \cite[Section 4.9]{Nee:B10} to show the stability of the queues.
\begin{lemma} \label{lemma:multi_slot_drift}
The multi-slot drift of interval $[t, t+T-1]$ ensures that:
\begin{align} \label{eq:multi_slot_drift}
\Delta_T (\V{Q}(t))
& \triangleq \frac{ \| \V{Q}(t+T-1) \|^2 - \| \V{Q}(t) \|^2 }{2} \nonumber \\
& \hspace{-15pt} \leq BT^2 - \sum_{e \in \Set{V}\,\cup\,\Set{E}} Q_e(t) \sum_{s=t}^{t+T-1} \frac{C_e - \tilde{a}_e^*(s)}{C_e}.
\end{align}
\end{lemma}

\begin{IEEEproof}
For each interface $e$, we can verify:
\begin{align}
\tilde{Q}_e(t+1) & \geq \tilde{Q}_e(t) - \big| C_e - \tilde{a}_e(t) \big| \geq \tilde{Q}_e(t) - \sqrt{2B_e}, \\
\tilde{Q}_e(t+1) & \leq \tilde{Q}_e(t) + \big| C_e - \tilde{a}_e(t) \big| \leq \tilde{Q}_e(t) + \sqrt{2B_e}.
\end{align}
in which we use the fact that $\big| C_e - \tilde{a}_e(t) \big| \leq \sqrt{2B_e}$. 
Iterating the queuing dynamics, we obtain:
\begin{align}
\hspace{-6pt} \tilde{Q}_e(t) - \sqrt{2B_e} (s - t) \leq \tilde{Q}_e(s) \leq \tilde{Q}_e(t) + \sqrt{2B_e} (s - t)
\end{align}
for any $s\geq t$, and furthermore
\begin{align} \begin{split}
& \hspace{13pt} (C_e - \tilde{a}_e^*(s)) \tilde{Q}_e(s) \\
& \leq (C_e - \tilde{a}_e^*(s)) \tilde{Q}_e(t) + \sqrt{2B_e}(s - t) \left| C_e - \tilde{a}_e^*(s) \right| \\
& \leq (C_e - \tilde{a}_e^*(s)) \tilde{Q}_e(t) + 2 B_e (s - t),
\end{split} \end{align}
therefore,
\begin{align} \begin{split}
& \hspace{12pt} \sum_{s=t}^{t+T-1} (C_e - \tilde{a}_e^*(s)) \tilde{Q}_e(s) \\
& \leq \sum_{s=t}^{t+T-1} (C_e - \tilde{a}_e^*(s)) \tilde{Q}_e(t) + 2B_e \sum_{s=t}^{t+T-1} (s - t) \\
& = \tilde{Q}_e(t) \sum_{s=t}^{t+T-1} (C_e - \tilde{a}_e^*(s)) + B_e \, T(T-1).
\end{split} \end{align}
Divide the result by $C_e^2$ and substitute it into the multi-slot drift, i.e., the sum of one-slot drifts \eqref{eq:one_slot_drift}, and we obtain:
\begin{align} \begin{split}
\Delta_T (\V{Q}(t))
& = \sum_{s=t}^{t+T-1} \Delta (\V{Q}(s)) \\
& \leq BT - \sum_{e \in \Set{V}\,\cup\,\Set{E}} \sum_{s = t}^{t+T-1} \frac{C_e - \tilde{a}_e^*(s)}{C_e} Q_e(s) \\
& \leq BT^2 - \sum_{e \in \Set{V}\,\cup\,\Set{E}} Q_e(t) \sum_{s=t}^{t+T-1} \frac{C_e - \tilde{a}_e^*(s)}{C_e}.
\end{split} \end{align}
which is \eqref{eq:multi_slot_drift}.
\end{IEEEproof}

Assume the state space (of the underlying Markov process) has a state ``0'' that we designate as a ``renewal'' state. Let sequence $\{ t_r : r \geq 0 \}$ represent the recurrence times to state 0, and $T_r = t_{r+1} - t_r$.
By renewal theory, we know that $T_r$ are i.i.d. random variables (and we denote its first and second moment by $\E{T}$ and $\E{T^2}$, respectively), and
\begin{align} \begin{split} 
& \hspace{12pt} \E{ \sum_{s=t_r}^{t_r + T_r - 1} \tilde{a}_e^*(s) \Bigg| \V{Q}(t)} = \E{ \sum_{s=0}^{T - 1} \tilde{a}_e^*(s) } \\
& = \sum_{c} \E{ \sum_{s = 0}^{T - 1} a^{(c)}(s) }  \sum_{\sigma \in \Set{F}_c(x)} \rho_e^{(\sigma)} \mathbb{P}_c(\sigma) \\
& = \E{T} \sum_{c} \lambda^{(c)} \sum_{\sigma \in \Set{F}_c(x)} \rho_e^{(\sigma)} \mathbb{P}_c(\sigma).
\end{split} \end{align}
Take (conditional) expectation of \eqref{eq:multi_slot_drift}, use the above result, and we can obtain \eqref{multi_slot_drift_2},
\begin{align} \begin{split} \label{multi_slot_drift_2}
& \hspace{12pt} \E{ \Delta_{T_r} (\V{Q}(t_r)) | \V{Q}(t_r) } \\
& \leq B \, \E{T^2} - \E{T} \sum_{e\in \Set{V}\,\cup\,\Set{E}}  \\
& \hspace{32pt}  \left( \frac{C_e - \sum_{c} \lambda^{(c)} \sum_{\sigma \in \Set{F}_c(x)} \rho_e^{(\sigma)} \mathbb{P}_c(\sigma)}{C_e}  \right) Q_e(t_r) \\
& \overset{\text{(c)}}{\leq} B \, \E{T^2} - \E{T} \epsilon \| \V{Q}(t_r) \|_1
\end{split} \end{align}
where we plug in \eqref{eq:slackness} to derive (c). Then take expectation with respect to $\V{Q}(t_r)$, and the result implies that the virtual queues are rate stable \cite[Theorem 4.12]{Nee:B10}.

\subsection{Stability of Actual Queues}
\label{apdx:actual_q}

We will show that the total backlog of actual queues, i.e.,
\begin{align} \label{eq:queue_component}
    R_{\operatorname{tot}}(t) = R'(t) + R(t)
\end{align}
is rate stable, where
\begin{align}
    R'(t) = \sum_{i\in \Set{V}} R_i'(t),\ 
    R(t) = \sum_{e\in \Set{V}\,\cup\,\Set{E}} R_e(t)
\end{align}
denote the backlogs of waiting queues and processing/ transmission queues.

\subsubsection{An Equivalent Problem}

We first derive an equivalent problem in this section.

Note that the waiting queue $R'(t)$ can be bounded by a linear function of the processing/transmission queue $R(t)$, i.e., 
\begin{align}
    R'(t) \leq \big( M_{\max} Z_{\max} \big) R(t) 
\end{align}
where $M_{\max} = \max_{\phi} M_{\phi}$, and $Z_{\max} = \max_{\phi} Z_{\phi}$ in which $Z_{\phi}$ denotes largest ratio of the sizes of any two packets associated with service $\phi$. The above inequality is valid because: for any packet staying in the waiting queue $R'(t)$, its associate must be in transit and thus collected in some processing/transmission queue, i.e., a component of $R(t)$. In the coefficient, factor $M_{\max}$ results from multi-step processing, i.e., there can be {\em multiple} packets associated with a given request staying in the waiting queue, and the number is bounded by $M_{\phi} - 1$ (in which case $M_{\phi} - 1$ static objects wait for one in-transit live packet) and thus $M_{\max}$; factor $Z_{\max}$ accounts for the difference in the data packet size of the waiting and in-transit packets. Substitute it into \eqref{eq:queue_component}, and $R_{\operatorname{tot}}(t)$ is also bounded by a linear function of $R(t)$, which implies that prove that $R_{\operatorname{tot}}(t)$ and $R(t)$ have the same stability property.

Furthermore, we define the {\em processing/transmission resource load} process at each network location, i.e.,
\begin{align}
    \tilde{R}_i(t) = \sum_{\psi \in \Set{R}_i(t)} r_{\psi} |\psi|,\ 
    \tilde{R}_{ij}(t) = R_{ij}(t)
\end{align}
where $\psi$ denotes a packet in the processing queue $\Set{R}_i(t)$, and $r_{\psi}$ the corresponding workload. Due to the linear relationship between $R(t)$ and $\tilde{R}(t)$, the remaining problem is to show that $\tilde{R}(t)$ is rate stable.

\subsubsection{Stability of $R(t)$}

We note that this part of proof is an extension to \cite[Appendix D]{SinMod:J18} (dealing with the simplified communication network setting).

Denote by $\tilde{R}^{(\kappa)}(t)$ the resource load incurred by all hop $\kappa$ packets (i.e., a packet that is $\kappa$ hops away from its source in the \ac{alg}), $\tilde{R}_e(t)$ the resource load at interface $e$, and $\tilde{R}_e^{(\kappa)}(t)$ the resource load at interface $e$ incurred by hop $\kappa$ packets.

For any interface $e$, and time $t_0$ and $t$, denote by $A_e(t_0, t)$ the additional resource load imposed on, and $S_e(t_0, t)$ the resource served at, interface $e$ during $[t_0, t]$; then,
\begin{align}
    A_e(t_0, t) \leq S_e(t_0, t) + o(t)
\end{align}
where $o(t)$ is a non-decreasing function, satisfying
\begin{align}
    \lim_{t\to\infty} \frac{o(t)}{t} = 0,
\end{align}
which is proved in \cite{zhang2021multicast}. Furthermore, let $\tilde{a}_e^{(\kappa)}(t_0, t)$ be the additional resource load, imposed by exogenously arriving requests during $[t_0, t]$, such that link $e$ is its $\kappa$-th hop in the assigned route. Note that
\begin{align}
    A_e(t_0, t) \geq \sum_{\kappa} \tilde{a}_e^{(\kappa)}(t_0, t)
\end{align}
because $A_e(t_0, t)$ includes packets already in the network before time $t_0$.

Then, we show the {\bf Proposition}: $\tilde{R}^{(\kappa)}(t) \leq B^{(\kappa)}(t)$ where $B^{(\kappa)}(t)$ is a non-decreasing function, satisfying $\lim_{t\to\infty} B^{(\kappa)}(t)/t = 0$. Assume empty queues at time $0$.

\underline{\em Base case:}
Let $t_0$ be the largest time when there is no hop $0$ packets waiting for operation at interface $e$.
Then
\begin{align} \begin{split}
\tilde{R}_e^{(0)}(t) & = \tilde{a}_e^{(0)}(t_0, t) - S_e(t_0, t) \\
& \leq A_e(t_0, t) - S_e(t_0, t) \leq o(t)
\end{split} \end{align}
and thus
\begin{align}
\tilde{R}^{(0)}(t) = \sum_{e\in \Set{V} \, \cup \, \Set{E}} R_e^{(0)}(t)
\leq (|\Set{V}| + |\Set{E}|) o(t)  = B^{(0)}(t),
\end{align}
and by assumption, the {\bf Proposition} holds.

\underline{\em Induction step:}
Assume $\tilde{R}^{(j)}(t) \leq B^{(j)}(t)$ where $B^{(j)}(t)$  is a non-decreasing function, satisfying $\lim_{t\to\infty} B^{(j)}(t)/t = 0$ for $j = 0,\cdots, \kappa - 1$. Let $t_0$ be the largest time when there is no hop $\kappa$ packets waiting for operation at interface $e$.

First, we focus on the resource load incurred by hop $\kappa$ packets arriving at $e$ during $[t_0, t]$, which are either (i) packets that were hop $0, \cdots, \kappa-1$ at time slot $t_0$, or (ii) exogenously arriving requests during $[t_0, t]$ that have $e$ as its $\kappa$-th hop. Therefore, the resource load incurred by hop $\kappa$ packets at interface $e$ is bounded by:
\begin{align}
    A_e^{(\kappa)}(t_0, t) \leq \sum_{j=0}^{\kappa-1} \tilde{\Xi}_{\max} B^{(j)}(t_0)
+ \tilde{a}_e^{(\kappa)}(t_0, t)
\end{align}
where $\tilde{\Xi}_{\max}$ is a constant factor handling the scaling of (i) data size due to processing, and (ii) processing/transmission resource units (because the edges in the ALG can be processing/transmission edges), given by
\begin{align}
    \tilde{\Xi}_{\max} = \Big( \max_{\phi, m_1, m_2} \prod_{m = m_1}^{m_2} \xi_{m}^{(\phi)} \Big) \Big( \max_{\phi, m} \Big\{ r_{m}^{(\phi)}, \frac{1}{r_{m}^{(\phi)}} \Big\} \Big).
\end{align}
 
Second, we consider the available resource to serve hop $\kappa$ packets, which is denoted by $S_e^{(\kappa)}(t_0, t)$ and satisfies:
\begin{subequations} \begin{align}
    S_e^{(\kappa)}(t_0, t) & \geq S_e(t_0, t) \\
    & \hspace{20pt} - \sum_{j=0}^{\kappa-1} \bigg[ \tilde{\Xi}_{\max} B^{(j)}(t_0)
    + \tilde{a}_e^{(j)}(t_0, t) \bigg] \label{eq:higher_priority}
\end{align} \end{subequations}
where \eqref{eq:higher_priority} is the resource consumed by packets of a higher priority (i.e., hop $0, \cdots, \kappa-1$ packets).

By definition, the remaining resource load incurred by hop $\kappa$ packets satisfies:
\begin{align} \label{eq:remaining_hop_k}
\tilde{R}_e^{(\kappa)}(t) & = A_e^{(\kappa)}(t_0, t) - S_e^{(\kappa)}(t_0, t) \nonumber \\
& \leq 2 \tilde{\Xi}_{\max} \sum_{j=0}^{\kappa-1} B^{(j)}(t_0) + \sum_{j=0}^{\kappa} \tilde{a}_e^{(j)}(t_0, t) - S_e(t_0, t) \nonumber \\
& \leq 2 \tilde{\Xi}_{\max} \sum_{j=0}^{\kappa-1} B^{(j)}(t) + A_e(t_0, t) - S_e(t_0, t) \nonumber \\
& \leq 2 \tilde{\Xi}_{\max} \sum_{j=0}^{\kappa-1} B^{(j)}(t) + o(t)
\end{align}
and thus
\begin{align}
\tilde{R}^{(\kappa)}(t) = \sum_{e\in \Set{V} \, \cup \, \Set{E}} \tilde{R}_e^{(\kappa)}(t) \leq B^{(\kappa)}(t)
\end{align}
in which
\begin{align}
B^{(\kappa)}(t) \triangleq  (|\Set{V}| + |\Set{E}|) \bigg[ 2 \tilde{\Xi}_{\max} \sum_{j=0}^{\kappa-1} B^{(j)}(t) + o(t) \bigg]
\end{align}
satisfying
\begin{align} \begin{split}
\lim_{t\to\infty} \frac{B^{(\kappa)}(t)}{t} = 0.
\end{split} \end{align}
Therefore, the {\bf Proposition} holds for case $\kappa$.

By induction, $\tilde{R}^{(\kappa)}(t)$ are stable for all $\kappa$, and so is the sum of them, i.e., $\tilde{R}(t)$, concluding the proof.

\section{Flow-based Characterization}
\label{apdx:nsr_flow}

We will show that an arrival vector $\V{\lambda}$ satisfies \eqref{eq:nsr} (referred to as {\em route-based} characterization) if and only if it satisfies \eqref{eq:resource_constraints} -- \eqref{eq:static_processed} (referred to as {\em flow-based} characterization).

\subsection{Necessity}

For any $\V{\lambda}$ satisfying \eqref{eq:nsr}, we define the flow variables in the \ac{alg} as follows (take live flows as an example):
\begin{subequations} \begin{align}
f_{i_m i_{m+1}}^{(c)}
& = \lambda^{(c)} \sum_{\sigma \in \Set{F}_{c}(x)} \frac{ w^{(c)}_{i_m i_{m+1}} }{ r_{m}^{(\phi)} }  \V{1}_{\{ (i_m, i_{m+1}) \in \sigma\}} \mathbb{P}_c{(\sigma)}, \\
f_{i_m j_m}^{(c)}
& = \lambda^{(c)} \sum_{\sigma \in \Set{F}_{c}(x)} w^{(c)}_{i_m j_m} \V{1}_{\{ (i_m, j_m) \in \sigma\}} \mathbb{P}_c{(\sigma)}.
\end{align} \end{subequations}
The goal is to show that these flow variables satisfy the flow-based characterization \eqref{eq:resource_constraints} -- \eqref{eq:static_processed}.

\subsubsection{Capacity Constraint}

We first verify the capacity constraints. Note that processing resource consumption at node $i$ is given by:
\begin{align} \begin{split}
& \hspace{2pt} \sum_{c, m} r_{m}^{(\phi)} f_{i_m i_{m+1}}^{(c)} \\
& \hspace{-10pt} = \sum_{c, m} \lambda^{(c)} \sum_{\sigma \in \Set{F}_{c}(x)} r_{m}^{(\phi)} \, \frac{w_{i_m i_{m+1}}^{(c)}}{ r_{m}^{(\phi)} } \, \V{1}_{\{ (i_m, i_{m+1}) \in \sigma\}} \mathbb{P}_c{(\sigma)} \\
& \hspace{-10pt} = \sum_{c} \lambda^{(c)} \sum_{\sigma \in \Set{F}_{c}(x)} \Big[ \sum_{m} w_{i_m i_{m+1}}^{(c)} \V{1}_{\{ (i_m, i_{m+1}) \in \sigma\}} \Big] \mathbb{P}_c{(\sigma)} \\
& \hspace{-10pt} = \sum_{c} \lambda^{(c)} \sum_{\sigma \in \Set{F}_{c}(x)} \rho_i^{(c, \sigma)} \mathbb{P}_c{(\sigma)}
\leq C_i,
\end{split} \end{align}
which satisfies the processing capacity constraint. The transmission resource constraint can be verified in a similar way and thus is omitted.

\subsubsection{Chaining Constraint}

Next, we verify the chaining constraint, taking an intermediate node $i_m$ (that is not a source or a destination) for example, and the goal is to show \eqref{eq:live_conservation}, i.e.,
\begin{align} \label{eq:local_flow_conservation}
& \hspace{12pt} \xi_{m}^{(\phi)} f_{i_{m-1} i_m}^{(c)} + \sum_{j\in \delta^{-}(i)} f_{j_m i_m}^{(c)} - f_{i_m i_{m+1}}^{(c)} - \sum_{j\in \delta^{+}(i)} f_{i_m j_m}^{(c)} \nonumber \\
& = \lambda^{(c)} \sum_{\sigma \in \Set{F}_{c}(x)} \mathbb{P}_c{(\sigma)} \times \Big[ \xi_{m}^{(\phi)} \frac{ w_{i_{m-1} i_m}^{(c)} }{ r_{m-1}^{(\phi)} } \, \V{1}_{\{ (i_{m-1}, i_m) \in \sigma\}} \nonumber \\
& \hspace{12pt} + \sum_{j\in \delta^{-}(i)} w_{j_m i_m}^{(c)} \V{1}_{\{ (j_m, i_m) \in \sigma\}} - \frac{ w^{(c)}_{i_m i_{m+1}} }{ r_{m}^{(\phi)} } \, \V{1}_{\{ (i_m, i_{m+1}) \in \sigma\}} \nonumber \\
&  \hspace{12pt} - \sum_{j\in \delta^{+}(i)} w_{i_m j_m}^{(c)} \V{1}_{\{ (i_m, j_m) \in \sigma\}} \Big]
= 0.
\end{align}
We will show that for each \ac{ER} $\sigma$, the square bracket equals $0$.
For each node $i_m$ in the \ac{alg}, one of the following two cases must be true:
\begin{enumerate}
    \item $i_m \notin \sigma$: all terms are equal to $0$, and \eqref{eq:local_flow_conservation} holds;
    \item $i_m \in \sigma$: the node must have exactly one incoming and one outgoing edge in the live data pipeline (i.e., one positive term and one negative), and by \eqref{eq:edge_load}:
    \begin{align*}
    & \xi_{m}^{(\phi)} w_{i_{m-1} i_m}^{(c)} / r_{m-1}^{(\phi)} ( = \xi_{m}^{(\phi)} \Xi_{m-1}^{(\phi)} ) = \Xi_{m}^{(\phi)}, \\
    & w_{i_m i_{m+1}}^{(c)} / r_{m}^{(\phi)} = \Xi_{m}^{(\phi)},\ w_{i_m j_m}^{(c)} = w_{j_m i_m}^{(c)} = \Xi_{m}^{(\phi)}.
    \end{align*}
    Therefore, in any case, the square bracket equals to $\Xi_{m}^{(\phi)} - \Xi_{m}^{(\phi)} = 0$, and \eqref{eq:local_flow_conservation} holds.
\end{enumerate}
To sum up, \eqref{eq:local_flow_conservation} holds for each intermediate node in the \ac{alg}. The conservation of live flow at source and destination nodes, as well as static flows, can be verified in a similar way and thus is omitted.

\subsection{Sufficiency}

Next, we show that for any flow variables $f, f'$ satisfying \eqref{eq:resource_constraints} -- \eqref{eq:static_processed}, there exists probability values associated with the possible \acp{ER} satisfying \eqref{eq:nsr}.

\subsubsection{Path-Finding for Stage $m$ Live Flow}

We divide the problem by (i) live and static pipelines, and (ii) by different processing stages. In particular, we focus on the stage $m \in \{ 1, \cdots, M \}$ live flow of client $c$.

Define a graph, whose node and edge sets are given by: $\Set{V} \cup \{ u, v \}$ and $\Set{E} \cup \{ (u, i): i\in \Set{V} \} \cup \{ (i, v): i\in \Set{V} \}$, where $\Set{V}$ and $\Set{E}$ denotes the node and link sets of the actual network, and $u$ and $v$ can be interpreted as the super source and destination. In addition, each edge is assigned the following capacity: $E_{ui} = \xi_{m-1}^{(\phi)} f_{i_{m-1} i_m}^{(c)}$, $E_{iv} = f_{i_m i_{m+1}}^{(c)}$, $E_{ij} = f_{i_m j_m}^{(c)}$.

Take $(u, v)$ as the source-destination pair, and the max-flow of the above graph is given by $\sum_{i\in \Set{V}} E_{ui}$, because: (i) $\{ E_{ij} \}$ is a set of feasible flow variables satisfying flow conservation constraint (imposed by the linear programming formulation of the max-flow \cite[Section 29.2]{cormen2009introduction}), and the resulting flow $\sum_{i\in \Set{V}} E_{ui}$ is an achievable value; (ii) while the max-flow is bounded by the total capacity of outgoing links from the source, which equals to the same value.

By running some standard max-flow algorithms, we can find a set of paths (from $u$ to $v$) achieving the max-flow, e.g., {\em Edmonds-Karp} algorithm, which finds the shortest augmented path \cite[Section 26.2]{cormen2009introduction} in each iteration (and thus is acyclic). We denote a path found by the algorithm by $p$, with associated rate (i.e., {\em bottleneck} of the augmenting path in the algorithm) denoted by $\beta_1^{(p)}$; besides, denote by $p(src)$ and $p(des)$ the first and last nodes other than $u, v$ in path $p$, representing its source and destination.

We can find paths for static flows by the same procedure, except that we use the super static source $o_m'$ instead of the super source introduced above. A found path is denoted by $q$ with an assigned rate of $\beta_2^{(q)}$.

\subsubsection{Composition of Individual Paths}

Next, we compose the separate paths into \acp{ER} and define the associated probability values for them.

\begin{figure}
    \centering
    \includegraphics[width = .99 \columnwidth]{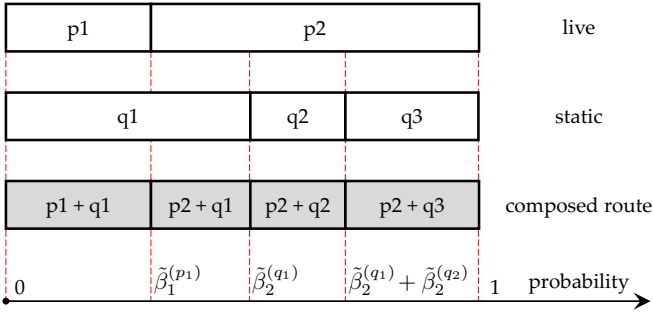}
    \caption{Composing live and static paths into routes, with the length of each segment equal to the corresponding probability value $\tilde{\beta}$.}
    \label{fig:composition}
\end{figure}

First, we compose the paths of live and static flows of the same stage. For each node $i$, denote the set of all incoming live and static paths by:
\begin{align}
\sigma_1^{(m, i)} = \big\{ p: p(des) = i \big\},\ 
\sigma_2^{(m, i)} = \big\{ q: q(des) = i \big\}
\end{align}
and the percentage of each path
\begin{align}
\tilde{\beta}_1^{(p)}
= \frac{ \beta_1^{(p)} }{ \sum\nolimits_{p'\in \sigma_1^{(m, i)}} \beta_1^{(p')} },\ 
\tilde{\beta}_2^{(q)}
= \frac{ \beta_2^{(q)} }{ \sum\nolimits_{q'\in \sigma_2^{(m, i)}} \beta_2^{(q')} }.
\end{align}
As illustrated in Fig. \ref{fig:composition}, we align the percentiles (i.e., cdfs) of the live and static paths, resulting in $|\sigma_1^{(m, i)}| + |\sigma_2^{(m, i)}| - 1$ non-overlapped segments. Each gray segment denotes a route composing the corresponding live and static paths. The routes and the percentage (i.e., length of the segments) are denoted by $z^{(m)}$ and $\tilde{\beta}^{(m)}$, and we define $z^{(m)}(src) = p^{(m)}(src)$ and $z^{(m)}(des) = p^{(m)}(des)$ where $p^{(m)}$ is the live path associated with $z^{(m)}$.

Then, we compose routes of different stages, say stage $m-1$ and $m$. For each node $i$, denote the set of all stage $m-1$ incoming routes and stage $m$ outgoing routes by 
\begin{subequations}
\label{eq:combined_route}
\begin{align} 
    \sigma^{(m-1, \, \to i)} & = \big\{ z^{(m-1)}: z^{(m-1)}(des) = i \big\} \\
    \sigma^{(m, \, i \to)} & = \big\{ z^{(m)}: z^{(m)}(src) = i \big\}
\end{align}
\end{subequations}
Similar to the procedure illustrated in Fig. \ref{fig:composition}, we align the percentiles of stage $m-1$ and $m$ routes, and each resulting segment represents a composed route of stage $m-1$ and $m$. Treat it as an aggregated stage, and repeat this procedure for other stages -- e.g., we can treat the routes of stage $m$ and $m+1$ as a single stage, and compose them with stage $m-1$ routes (to generate routes of stage $m-1$ to $m+1$) -- until routes of all stages are obtained. The results are a set of \acp{ER} $\{ \sigma \}$ and the associated probability values $\{ \mathbb{P}_c^{(\sigma)} \}$ that equal to the length of corresponding segments.

It is clear that no additional resource loads are imposed on any network location, and thus the incurred resource consumption (which can be represented by $\{ \mathbb{P}_c^{(\sigma)} \}$) does not violate the capacity constraints \eqref{eq:resource_constraints}, leading to \eqref{eq:nsr}.

\section{Stability Region with Dynamic Replacement}
\label{apdx:nsr_replacement}

\subsection{Necessity}

Consider an arrival process $a(t) = \{ a^{(c)}(t) : \forall\, c \}$ and the policy able to support it, i.e., keeping the actual queues stable. Fix a time interval $[0, t-1]$, and denote by $x(\tau)$ the database placement at time $\tau$, $\Set{T}(x)$ the set of time slots with database placement $x$, and $a^{(c, \sigma)}(\tau)$ the number of requests of client $c$ emerging at time $\tau$ and successfully delivered by the end of the interval (i.e., time $t-1$).
Define
\begin{align} \label{eq:x_prob}
\mathbb{P}(x) \triangleq \lim_{t\to\infty} \frac{|\Set{T}(x)|}{t} \ \Rightarrow\ 
\sum_{x\in \Set{X}} \mathbb{P}(x) = 1,
\end{align}
and without loss of generality, we assume $\mathbb{P}(x) > 0$ (otherwise, we can exclude $x$ from $\Set{X}$).

By the conservation of requests of client $c$ emerging at time $\tau \in \Set{T}(x)$:
\begin{align} \label{eq:data_conservation2}
Y^{(c, x)}(t) + \sum_{\tau \in \Set{T}(x)} \sum_{\sigma \in \Set{F}_c(x)} a^{(c, \sigma)}(\tau) = \sum_{\tau \in \Set{T}(x)} a^{(c)}(\tau)
\end{align}
where $Y^{(c, x)}(t)$ is the amount of such packets staying in actual queues at time $t$.
Divide it by $|\Set{T}(x)|$, let $t\to\infty$:
\begin{align} \begin{split}
    & \lim_{t \to \infty} \frac{Y^{(c)}(t)}{|\Set{T}(x)|} = \lim_{t \to \infty} \frac{t}{|\Set{T}(x)|} \times \lim_{t \to \infty} \frac{Y^{(c)}(t)}{t} = 0, \\
    & \lim_{t \to \infty} \frac{1}{|\Set{T}(x)|} \sum_{\tau \in \Set{T}(x)} a^{(c)}(\tau) = \lambda^{(c)}.
\end{split} \end{align}
In addition, define
\begin{align} \begin{split}
    \lambda^{(\sigma; c, x)} & \triangleq \lim_{t \to \infty} \frac{1}{|\Set{T}(x)|} \sum_{\tau \in \Set{T}(x)} a^{(c, \sigma)}(\tau), \\ 
    \mathbb{P}_{c, x}(\sigma) & \triangleq \frac{\lambda^{(\sigma; c, x)} }{\lambda^{(c)} }, \label{eq:xc_prob}
\end{split} \end{align}
and \eqref{eq:data_conservation2} becomes
\begin{align} \label{eq:split_by_x}
\sum_{\sigma \in \Set{F}_c(x)} \mathbb{P}_{c, x}(\sigma) = 1,\ \forall\, c.
\end{align}

On the other hand, the total resource consumption (up to time $t$) at each interface $e \in \Set{V} \cup \Set{E}$ satisfies:
\begin{align}
C_e \, t & \geq \sum_{\tau = 0}^{t-1} \sum_{x \in \Set{X}} \V{1}_{\{x(\tau) = x\}}  \sum_{c} \sum_{\sigma \in \Set{F}_c(x)} \rho_e^{(c, \sigma)} a^{(c, \sigma)}(\tau) \nonumber \\
& = \sum_{x \in \Set{X}} \sum_{c} \sum_{\sigma \in \Set{F}_c(x)} \rho_e^{(c, \sigma)} \sum_{\tau\in \Set{T}(x)} a^{(c, \sigma)}(\tau).
\end{align}
Divide by $t$, let $t\to\infty$:
\begin{subequations} \begin{align}
C_e & \geq \sum_{x \in \Set{X}} \lim_{t\to\infty} \frac{|\Set{T}(x)|}{t}
\sum_{c} \sum_{\sigma \in \Set{F}_c(x)} \rho_e^{(c, \sigma)} \times \nonumber \\
& \hspace{80pt} \lim_{T\to\infty} \frac{1}{|\Set{T}(x)|} \sum_{\tau\in \Set{T}(x)} a^{(c, \sigma)}(\tau) \\
& = \sum_{x \in \Set{X}} \mathbb{P}(x)
\sum_{c} \sum_{\sigma \in \Set{F}_c(x)} \rho_e^{(c, \sigma)} \lambda^{(\sigma; c, x)} \\
& = \sum_{x \in \Set{X}} \mathbb{P}(x)
\sum_{c} \lambda^{(c)} \sum_{\sigma \in \Set{F}_c(x)} \rho_e^{(c, \sigma)} \mathbb{P}_{c, x}(\sigma).
\end{align} \end{subequations}
which is \eqref{eq:nsr_dynamic}, concluding the proof.

\subsection{Sufficiency}

Suppose the arrival vector $\V{\lambda}$ satisfies \eqref{eq:nsr_dynamic}, i.e., there exists probability values $\mathbb{P}(x)$, $\mathbb{P}_{c, x}(\sigma)$ such that:
\begin{align} \label{eq:capacity_replace}
    C_e - \sum_{x \in \Set{X}} \mathbb{P}(x) \sum_{c} \lambda^{(c)} \sum_{\sigma \in \Set{F}_c(x)} \rho_e^{(c, \sigma)} \mathbb{P}_{c, x}(\sigma) \geq \epsilon C_e
\end{align}
for each interface $e$, and we will define a randomized route selection policy to keep the virtual queues stable.\footnote{
    Then following the same procedure as presented in Appendix \ref{apdx:actual_q}, we can show that: under the policy combining the designed route selection policy and ENTO, the actual queues are rate-stable.
}

\subsubsection{Initial Design}

We first assume that: ``each node can complete arbitrary database replacement immediately'', i.e., the caching vector $x(t) \in \Set{X}$ can be different at each time slot, and we define a reference policy * operating as follows:
at each time slot,
(i) select the database placement $x$ w.p. $\mathbb{P}(x)$,
(ii) select the \ac{ER} $\sigma$ to deliver the requests of client $c$ w.p. $\mathbb{P}_{c, x}(\sigma)$.

Following the procedure in Appendix \ref{apdx:virtual_stability}, the normalized virtual queue drift is given by:
\begin{align} \label{eq:drift_replacement} 
\E{ \Delta(Q_e(t)) }
\leq B_e' - \frac{ C_e - \E{ \tilde{a}_e^*(t) } }{C_e} \, \E{ Q_e(t) }
\end{align}
with
\begin{align}
B_e' = \frac{1}{2C_e^2} \max \left\{ C_e, \, \max_{x \in \Set{X}} \sum_{c} A_{\max}^{(c)} \max_{\sigma \in \Set{F}_c(x)} \rho_e^{(c, \sigma)} \right\}^2
\end{align}
satisfying $\big| C_e - \tilde{a}_e(t) \big| \leq \sqrt{2B_e'}$, and
\begin{align} \begin{split} \label{eq:load_replacement} 
\E{ \tilde{a}_e^*(t) } & = \E{\E{ \tilde{a}_e^*(t) | x }}  \\
& = \sum_{x \in \Set{X}} \mathbb{P}(x) \sum_{c} \lambda^{(c)} \sum_{\sigma \in \Set{F}_c(x)} \rho_e^{(c, \sigma)} \mathbb{P}_{c, x}(\sigma).
\end{split} \end{align}
Substitute \eqref{eq:load_replacement}, together with \eqref{eq:capacity_replace}, into \eqref{eq:drift_replacement}, and we obtain
\begin{align}
\E{ \Delta( Q_e(t) ) } \leq B_e' - \epsilon \, \E{Q_e(t)},
\end{align}
which implies that the virtual queue of each interface $e$ is rate-stable \cite[Section 3.1.4]{Nee:B10}.

\subsubsection{Low-Cost Replacement} \label{apdx:low-rate-replacement}

In this section, we aim to design a policy in a realistic setting assuming restricted transmission rate for database replacement.

Consider a two-timescale system, where processing and transmission decisions are made on a per time slot basis, and database replacement decisions are made on a per time frame basis, with each frame including $T$ consecutive slots.
Define a reference policy * operating as follows:
(i) at the beginning of each frame, select the database placement $x$ w.p. $\mathbb{P}(x)$,
(ii) at each time slot, select the \ac{ER} $\sigma$ to deliver the requests of client $c$ w.p. $\mathbb{P}_{c, x}(\sigma)$.

Let sequence $\{ rT : r\geq 0 \}$ represent the starting time slots of the time frames. According to Lemma \ref{lemma:multi_slot_drift}, the multi-slot drift of interval $[rT, (r+1)T - 1]$ is give by:
\begin{align}
\E{\Delta_T(Q_e(rT))} \leq B_e' T^2 - \epsilon T \, \E{Q_e(rT)}.
\end{align}
Apply the telescope sum \cite{Nee:B10} for $r \in [0, R-1]$:
\begin{align}
R B_e' T^2 - \epsilon T \sum_{r=0}^{R-1}\E{Q_e(rT)} \geq Q_e^2(RT) \geq 0.
\end{align}
and therefore, for any point interior to the stability region, i.e., $\epsilon > 0$,
\begin{align} \label{eq:point_q}
\sum_{r=0}^{R-1} \E{Q_e(rT)} \leq \frac{R B_e' T}{\epsilon}.
\end{align}
In addition, we note that:
\begin{align} \begin{split} \label{eq:frame_q}
\sum_{s = 0}^{T-1} \E{Q_e(rT + s)}
& \leq \sum_{s = 0}^{T-1} \big[ \E{Q_e(rT)} + \sqrt{2B_e'} s \big] \\
& \leq T\, \E{Q_e(rT)} + \frac{\sqrt{2B_e'} \, T^2}{2}.
\end{split}  \end{align}
Therefore, the average virtual queue backlog over interval $[0, t-1]$, with $t = RT + \Delta t\ (0\leq \Delta t < T)$, is given by
\begin{align}
    & \hspace{12pt} \frac{1}{t}\sum_{\tau = 0}^{t-1} \E{Q_e(\tau)} 
    = \frac{1}{RT + \Delta t}\sum_{\tau = 0}^{RT + \Delta t - 1} \E{Q_e(\tau)} \\
    & \leq \frac{1}{RT} \sum_{\tau = 0}^{(R+1)T - 1} \E{Q_e(\tau)}
    = \frac{1}{RT} \sum_{r = 0}^{R} \sum_{s=0}^{T-1} \E{Q_e(rT + s)} \nonumber \\
    & \overset{\text{(a)}}{\leq} \frac{1}{RT} \sum_{r = 0}^{R} \Big(  T\, \E{Q_e(rT)} + \frac{\sqrt{2B_e'} \, T^2}{2} \Big) \nonumber \\
    & \overset{\text{(b)}}{\leq} \frac{R+1}{R} \Big(  \frac{B_e'}{\epsilon} + \frac{\sqrt{2B_e'}}{2} \Big) T
\end{align}
where (a) and (b) result from \eqref{eq:frame_q} and \eqref{eq:point_q}, respectively. Let $t\to\infty$:
\begin{align}
    \lim_{t\to\infty} \frac{1}{t}\sum_{\tau = 0}^{t-1} \E{Q_e(\tau)} 
    & \leq \lim_{R\to\infty}  \frac{R+1}{R} \Big(  \frac{B_e'}{\epsilon} + \frac{\sqrt{2B_e'}}{2} \Big) T \nonumber \\
    & = \Big(  \frac{B_e'}{\epsilon} + \frac{\sqrt{2B_e'}}{2} \Big) T \sim \mathcal{O}(T),
\end{align}
i.e., the virtual queues are mean rate stable.

On the other hand, policy * incurs a cost of
\begin{align}
    \text{Replacement rate} \leq \frac{|\Set{V}| \sum_{k\in \Set{K}} F_k}{T} \sim \mathcal{O}(1/T).
\end{align}

To sum up, policy * achieves an $[\mathcal{O}(T), \mathcal{O}(1/T)]$ tradeoff between average (virtual) queue backlog and replacement rate.
While the policy remains throughput-optimal regardless of $T$, the attained replacement rate can be arbitrarily close to zero by pushing $T\to \infty$, with a tradeoff in the queue backlog (and thus the delay performance).

\section{Equivalence of the MILP Formulation}
\label{apdx:milp_equivalence}

Fix the database placement $x$, and we will show that the remaining problems of \eqref{eq:mip_placement} and \eqref{eq:milp_placement} are equivalent.

{\bf Suppose $x_{i, k} = 1$}, and we note that
\begin{align}
    \sum_{ j\in \delta^+ (i) } f_{ij}'^{(k)} + f_i'^{(k)} \leq C^{\max}_{i, k}
\end{align}
since the two terms on the left hand side are bounded by corresponding components of $C^{\max}_{i, k}$ \eqref{eq:C_max}, given that the flow variables satisfy the capacity constraints \eqref{eq:resource_constraints}.
Therefore, no additional constraints is imposed on the flow variables by \eqref{eq:static_conservation2}, equivalent to \eqref{eq:static_conservation}.

{\bf Suppose $x_{i, k} = 0$}, in which case \eqref{eq:static_conservation2} transforms to
\begin{align*}
\sum_{ j\in \delta^+ (i) } f_{ij}'^{(k)} + f_i'^{(k)} \leq \sum_{ j \in \delta^-(i) } f_{ji}'^{(k)}
\end{align*}
and it holds with equality for {\em any} solution satisfying \eqref{eq:static_conservation}.
Reversely, consider any flow assignment satisfying  \eqref{eq:static_conservation2} with inequality, i.e., the total incoming flow is strictly greater than the outgoing flow, which essentially results from the excessive static flow produced by the static sources. Then, we can reduce the incoming flow rate $f_{ji}'^{(k)}$ to make it an equality at node $i$, and repeat the same procedure for all neighbor nodes $j \in \delta_i^-$ impacted by the change, which terminates upon meeting the static sources $j \in \Set{V}(k_m^{(\phi)})$.


\end{document}